\documentclass[epj]{svjour}
\pdfoutput=1

\usepackage{graphicx}
\usepackage{dcolumn}
\usepackage{bm}

\usepackage{subfig}

\usepackage{array} 
\usepackage{lineno}
\usepackage{amsmath} 
\usepackage[numbers]{natbib}
\usepackage{comment}
\usepackage{stackrel}
\usepackage{mathtools}
\usepackage{multirow}
\usepackage{tabularx}
\usepackage{array}
\usepackage{hyperref}
\usepackage[thinlines]{easytable}

\allowdisplaybreaks

\begin{document}

\title{Probabilistic treatment of the uncertainty from the finite size of weighted Monte Carlo data}
\author{Thorsten Gl\"usenkamp
	\thanks{thorsten.gluesenkamp@fau.de}%
}                     
\institute{Erlangen Centre for Astroparticle Physics (ECAP), Erlangen}

\date{Received: date / Revised version: date}
%
\abstract{Parameter estimation in HEP experiments often involves Monte-Carlo simulation to model the experimental response function. A typical application are forward-folding likelihood analyses with re-weighting, or time-consuming minimization schemes with a new simulation set for each parameter value. Problematically, the finite size of such Monte Carlo samples carries intrinsic uncertainty that can lead to a substantial bias in parameter estimation if it is neglected and the sample size is small. We introduce a probabilistic treatment of this problem by replacing the usual likelihood functions with novel generalized probability distributions that incorporate the finite statistics via suitable marginalization. These new PDFs are analytic, and can be used to replace the Poisson, multinomial, and sample-based unbinned likelihoods, which covers many use cases in high-energy physics. In the limit of infinite statistics, they reduce to the respective standard probability distributions. In the general case of arbitrary Monte Carlo weights, the expressions involve the fourth Lauricella function $F_D$, for which we find a new finite-sum representation in a certain parameter setting. The result also represents an exact form for Carlson's Dirichlet average $R_n$ with $n>0$, and thereby an efficient way to calculate the probability generating function of the Dirichlet-multinomial distribution, the extended divided difference of a monomial, or arbitrary moments of univariate B-splines. We demonstrate the bias reduction of our  approach with a typical toy Monte Carlo problem, estimating the normalization of a peak in a falling energy spectrum, and compare the results with previously published methods from the literature.
} 
\maketitle

\tableofcontents 

\section{Introduction}
\subsection{Monte Carlo-based parameter estimation}

Parameter estimation, or inference of parameters from the measured data, is the central goal of modern experiments. In this setting, the likelihood function plays the role of probabilistically connecting the data to the parameters of interest, in both Frequentist and Bayesian applications \cite{Lyons2013}. An implicit component required to calculate the likelihood function in an experiment is the experimental response, i.e. how some "true" and measured quantities are connected to each other via the experimental setup. A "true" quantity could for example be the a priori unknown direction or energy of a particle entering a particle detector (e.g. in a neutrino detector like IceCube \cite{Halzen2006} or in an experiment at the Large Hadron Collider \cite{Evans2008}), while the measured quantity denotes any estimator of the same. Since the experimental response is a complicated function that often cannot be written down analytically, it is usually approximated by Monte Carlo (MC) simulations \cite{Binder2010}. 

The MC events are typically binned in histograms to approximate the distributions in the desired variables. This is useful, since binning ensures an exact control of the underlying statistics\footnote{Sometimes the histogram is smoothed, for example by spline interpolation. This introduces extra uncertainty which is not straightforward to quantify. We will not deal with this issue here.}. These histograms are then used to calculate the likelihood function itself. If normalization plays a role, this is usually done in the form of individual Poisson factors ("Poisson Likelihood" $L_\mathrm{\mathbf{P}}$),

\begin{equation}
L_{\mathrm{\mathbf{P}}}(\theta)=\prod_{\mathrm{bins} \ i} \frac{{e^{-\lambda_i(\theta)}}{\lambda_i(\theta)}^{k_i} }{k_i!} \label{eq:poisson_1}=\frac{{e^{-\lambda(\theta)}}{\lambda(\theta)}^{k} }{k!} \cdot L_{\mathrm{\mathbf{MN}}}(\theta)  
\end{equation}
, where $\lambda_i, k_i$ ($\lambda, k$) denote the expectation value and observed number of events in bin $i$ (in all bins), and $\theta$ stands for the parameters to be inferred. The observed $k_i$ are assumed to be independent. $L_{\mathrm{\mathbf{MN}}}$ denotes the multinomial likelihood, which is connected to the Poisson likelihood via a global Poisson factor that encompasses all events in all bins. The multinomial likelihood is sometimes implicitly used to approximate a PDF (probability density function) via a histogram for unbinned likelihood approaches as

\begin{align}
L_{\mathrm{\mathbf{MN}}}(\theta)
&=k! \cdot \prod_{\mathrm{bins} \ i} \frac{1}{{k_i}!} \left(\frac{\lambda_i(\theta)}{\lambda(\theta)}\right)^{k_i}
=k! \cdot\prod_{\mathrm{bins} \ i} \frac{ {p_i(\theta)}^{k_i} }{k_i!} \\
&= k! \cdot \left( \prod_{\mathrm{bins} \ i} \frac{\mathrm{vol}_{\mathrm{bin,i}}^{k_i}}{k_i!} \right) \cdot \prod_{\mathrm{evs} \ j}  \overbrace{\frac{p_{i(j)}(\theta)}{\mathrm{vol}_{\mathrm{bin,i(j)}}}   }^{f_{\mathrm{approx.},\theta}(x_j)}
= K \cdot \prod_{\mathrm{evs} \ j} f_{\mathrm{approx.},\theta}(x_j) \\ 
&= K \cdot L_{\mathrm{\textbf{U}}}(\theta)
\end{align}
, where $p_i$ is the relative probability of an event to lie in bin $i$, $\mathrm{vol}_\mathrm{bin,i}$ is the bin-width of bin $i$ or its higher-dimensional analogue, $p_{i(j)}=p_{i}$ and $\mathrm{vol}_{\mathrm{bin,i(j)}}=\mathrm{vol}_{\mathrm{bin,i}}$ for the corresponding bin $i$ which contains event $j$, $K=k! \cdot \left( \prod_{i} \frac{\mathrm{vol}_{\mathrm{bin,i}}^{k_i}}{k_i!} \right)$ is a proportionality constant, and $f_{\mathrm{approx.}}$ can be seen as a MC-based approximate PDF in the corresponding continuous observable $x$ used for an unbinned likelihood $L_{\mathrm{\mathbf{U}}}$. Since the approximate unbinned likelihood multiplied by the constant factor $K$ is equal to the multinomial likelihood, both yield the same results for parameter estimation. One can also write the approximate unbinned likelihood in terms of a product of multinomial factors with total count $1$, namely categorical distributions, via
\begin{align}
L_{\mathrm{\mathbf{U}}}=\prod_{\mathrm{evs} \ j}  \frac{p_{i(j)}(\theta)}{\mathrm{vol}_{\mathrm{bin,i(j)}}}=\prod_{\mathrm{evs} \ j} \frac{L_{\mathrm{\mathbf{MN}}}(k=1;\theta)_{i(j)}}{\mathrm{vol}_{\mathrm{bin,i(j)}}} = \prod_{\mathrm{evs} \ j} \frac{L_{\mathrm{\mathbf{Cat}}}(\theta)_{i(j)}}{\mathrm{vol}_{\mathrm{bin,i(j)}}} \label{eq:mnomial_unbinned_relation}
\end{align}
which is a form that makes it explicit that individual events are independent from each other. The term $i(j)$ again denotes the bin $i$ which contains event $j$. It can be seen that each categorical factor calculates the discrete probability $p_i$ of a given bin. This second relation for $L_{\mathrm{\mathbf{U}}}$ is relevant for the rest of the paper, because it is straightforward to generalize with the extended multinomial distributions discussed in section \ref{sec:multinomial}. 

The first part of the paper will focus on the Poisson likelihood $L_{\mathrm{\mathbf{P}}}$. The second part will cover the multinomial case $L_{\mathrm{\mathbf{MN}}}$, and thereby implicitly also the unbinned likelihood $L_{\mathrm{\mathbf{U}}}$. The last part discusses a toy-MC application with comparisons to some other approaches in the literature.

 \subsection{The problem: finite number of MC events}
 \label{sec:the_problem_definition}
 The crucial point in MC-based parameter estimation is the possibility to re-weight individual MC events based on a weighting function that depends on some parameters $\theta$ and is usually defined over the unobserved "true" space. A change of $\theta$ leads to change in the weighting function which leads to a changing bin content $\sum_{i} w_i$ in the histograms of the measurable quantities. 
 Therefore, the generic Poisson likelihood (eq. \ref{eq:poisson_1}) actually reads 
 
\begin{equation}
L_{\mathrm{\mathbf{P}}}(\theta)=\prod_{\mathrm{bins} \ i} \frac{{e^{-{\sum_j w_{i,j}(\theta)}}}{({\sum_j w_{i,j}(\theta)}})^{k_i} }{k_i!} \label{eq:poisson_with_sumweights}
\end{equation}
, i.e. the expectation value $\lambda_i$ in each bin $i$ is really given by the sum of weights in the given bin which itself depends on $\theta$. However, based on the amount of MC events that end up in a given bin, this approximation can be very poor. If only one MC event ends up in a bin, for example, the uncertainty of the true expectation value $\lambda$ is large, which can lead to a bias in a statistical analysis if it is taken to be exact, as in eq. (\ref{eq:poisson_with_sumweights}). Several papers in the literature have addressed this issue in the past: with additional minimization schemes for unknown "true" rates assuming average weights per bin \cite{Barlow1993} or exploiting the full weight distribution \cite{Chirkin2013}, for Gaussian approximations \cite{Bohm2012}, or using numerical techniques \cite{Bohm2014} in order to approximate a compound Poisson distribution, where the sum of weight random variables itself is Poisson distributed. 

In this paper, we describe an Ansatz that can be interpreted as a probabilistic counterpart to  \cite{Barlow1993} or \cite{Chirkin2013}, and also as an analytic approximation to the compound Poisson distribution advocated for in \cite{Bohm2014}, with the difference that the weight distributions are inferred on a per-weight basis and that the number of summands of random variables is fixed. Probabilistic approaches involve Priors, which might be the reason they have not really been studied in the past in this context. However, as we will show, the Prior distribution for this particular problem is not arbitrary, but can be fixed by knowledge about the problem at hand - namely having weighted MC events in a bin. The resulting expressions are generalizations of the Poisson ($L_{\mathrm{\mathbf{P}}}$), multinomial ($L_{\textrm{\textbf{MN}}}$) and approximated unbinned likelihood ($L_{\textrm{\textbf{U}}}$), and they are equal to the respective standard likelihood form in the limit of infinite statistics.

To sketch the generalization we derive for the Poisson likelihood, let us re-write eq. (\ref{eq:poisson_with_sumweights}) as a special case of an expectation value,

\begin{align}
L_{\mathrm{\mathbf{P}}}(\theta)&=\prod_{\mathrm{bins} \ i} \frac{{e^{-{\sum_j w_{j,i}(\theta)}}}{({\sum_j w_{j,i}(\theta)}})^{k_i} }{k_i!} \nonumber \\
&=\prod_{\mathrm{bins} \ i} \int_{0}^{\infty} \frac{{e^{-\lambda_i}}{\lambda_i}^{k_i} }{k_i!} \cdot  \delta\left(\lambda_i - {\sum_j^{n_i} w_{j,i}(\theta)}\right) d\lambda_i \label{eq:delta_assumption_for_base_formula} \\ &=  \prod_{\mathrm{bins} \ i} \int_{0}^{\infty} \frac{{e^{-\lambda_i}}{\lambda_i}^{k_i} }{k_i!} \cdot  \left[\delta(\lambda_{i,1} - w_{1,i}) \ast \ldots \ast \delta(\lambda_{i,n_i} - w_{n_i ,i})\right](\lambda_i) \ d\lambda_i \label{eq:base_formula_with_individual_deltas}
\end{align}
, where we first write the likelihood as an integral over the delta distribution (eq. \ref{eq:delta_assumption_for_base_formula}), and then rewrite the delta distribution as a convolution of $n_i$ individual delta distributions (eq. \ref{eq:base_formula_with_individual_deltas}), one for each weight $w_{j,i}$. So far, nothing has changed with respect to the standard Poisson likelihood. We now simply generalize this formula via 

\begin{align}
L_{\mathrm{\mathbf{P}, finite}}(\theta)&=\prod_{\mathrm{bins} \ i} \int_{0}^{\infty} \frac{{e^{-\lambda_i}}{\lambda_i}^{k_i} }{k_i!} \cdot \left[\mathrm{\mathbf{G}}(\lambda_{i,1};w_{1,i}) \ast \ldots \ast \mathrm{\mathbf{G}}(\lambda_{i,n_i};w_{n_i,i})\right](\lambda_i) \ d\lambda_i \label{eq:base_formula_with_convolution_of_gammas} \\
&=\prod_{\mathrm{bins} \ i} \int_{0}^{\infty} \frac{{e^{-\lambda_i}}{\lambda_i}^{k_i} }{k_i!} \cdot  P(\lambda_i, \bm{w_i}(\theta)) \ d\lambda_i \nonumber \\
&=\prod_{\mathrm{bins} \ i}\mathrm{E}\left[\frac{{e^{-\lambda_i}}{\lambda_i}^{k_i} }{k_i!}\right]_{P(\lambda_i)} = \prod_{\mathrm{bins} \ i}^{} L_{\mathrm{\mathbf{P}, finite, i}}(k_i; \bm{w_i}(\theta)) 
\label{eq:poisson_to_expectation_value}
\end{align}
, replacing each delta distribution by another distribution $\mathrm{\mathbf{G}}(\lambda_{i,j};w_{j,i})$, which depends on the weight as a scaling parameter, but does not possess infinite sharpness. The convolution of individual distributions is then written as $P(\lambda_i, \bm{w_i}(\theta))$, and the end result can be also interpreted as the expectation value of the Poisson distribution under a "more suited" distribution $P$, that is not just a delta distribution as in eq. (\ref{eq:delta_assumption_for_base_formula}). In the next section we will show that $\mathrm{\mathbf{G}}$ naturally can be identified with the gamma distribution, and has the interpretation of the inferred expectation value given a single MC event with weight $w_{j,i}$. The convolution of several such distributions is thus the natural probabilistic generalization from a sum of weights to a sum of random variables, where each random variable corresponds to a given weight.

It should be said that this Ansatz captures only a part of the total uncertainty if the number of MC events is not fixed, but for example follows a Poisson distribution. In this case, for given Monte Carlo parameters $\theta_0$ used during the generation, each realization will give a different Poisson-distributed number of Monte Carlo events $k_{mc,i}$ (previously denoted by $n_i$), which in turn will all have different weights each time if a re-weighting step with new parameters $\theta$ is applied. Therefore, one should in principle also integrate over another distribution $P_{\mathrm{samp.}}(k_{mc,i},\bm{w_i}; \theta, \theta_0)$, to take this additional uncertainty from the sampling step into account. The following few equations show that these two uncertainties can be treated separately. The total Ansatz looks like

\begin{align}
&\begin{aligned}
L_{\mathrm{\mathbf{P}, finite}}(\theta) &=
 \prod_{ \stackrel{\mathrm{bins}}{i}}   \sum \limits_{k_{mc,i}=0}^{\infty} \idotsint \limits_{\bm{w_i}}^{} \int \limits_{\lambda_i} \frac{{e^{-\lambda_i}}{\lambda_i}^{k_i} }{k_i!} \cdot  P(\lambda_i;k_{mc,i}, \bm{w_i})  \cdot P_{\mathrm{samp.}}(k_{mc,i},\bm{w_i}; \theta) \ d\lambda_i\bm{dw_i} 
\end{aligned} \\
&\begin{aligned}
=\prod_{ \stackrel{\mathrm{bins}}{i}}   \sum \limits_{k_{mc,i}=0}^{\infty} \idotsint \limits_{\bm{w_i}}^{} \int \limits_{\lambda_i} \frac{{e^{-\lambda_i}}{\lambda_i}^{k_i} }{k_i!}  \cdot  P(\lambda_i;k_{mc,i}, \bm{w_i}) \ d\lambda_i \cdot P_2(k_{mc,i}; \theta_0) \cdot P_3(\bm{w_i}; \theta) \ \bm{dw_i}  \label{eq:product_of_mcevents_and_weights}
\end{aligned} \\
&\begin{aligned}
&=\prod_{ \stackrel{\mathrm{bins}}{i}}   \sum \limits_{k_{mc,i}=0}^{\infty} \idotsint \limits_{\bm{w_i}}^{} L_{\mathrm{\mathbf{P}, finite, i}}(k_i; \bm{w_i}) \cdot P_2(k_{mc,i}; \theta_0) \cdot P_3(\bm{w_i}; \theta) \ \bm{dw_i}
\end{aligned} \\
&\begin{aligned}
&=\prod_{ \stackrel{\mathrm{bins}}{i}}  \idotsint \limits_{\bm{w_i}}^{} L_{\mathrm{\mathbf{P}, finite, i}}(k_i; \bm{w_i}) \cdot \delta(\bm{w_i}-\bm{w_i}(\theta)) \ \bm{dw_i} =\prod_{\mathrm{bins} \ i} L_{\mathrm{\mathbf{P}, finite, i}}(k_i; \bm{w_i}(\theta))  \label{eq:delta_assumption}
 \end{aligned}
\end{align}
where the discrete summation sums over all possible Monte Carlo sample outcomes, i.e. $k_{mc,i}$ denotes the number of MC events in bin $i$, and for each such count there is an integration over the respective weight distribution. In the process, we isolate the integration over $ \lambda_i$ and further make two simplifications. First, we assume we can split $P_{\mathrm{samp.}}(\bm{w_i}, k_{mc,i}; \theta, \theta_0)=P_2(k_{mc,i}; \theta_0) \cdot P_3(\bm{w_i}; \theta)$ in eq. (\ref{eq:product_of_mcevents_and_weights}), which seems reasonable since usually the total number of events $P_3$ depends on the overscaling factor or simulated live time (here implicit in the generation parameters $\theta_0$) and is independent of the weight distribution $P_3$ for a given $\theta$. The second simplification (eq. \ref{eq:delta_assumption}) neglects the actual sampling uncertainty, i.e. it only picks out the term in the summation corresponding to the actual simulated $k_{mc,i}$ and replaces $P_3$ with a delta function. 
For the rest of the paper, we work with this simplification, since the distribution over weights is usually intractable and its form depends on the given application. Notice, however, that whether we make this simplification or not, the integral over $\lambda_i$ can always be performed. 

During the writing of this paper we became aware of two other publications \cite{Beaujean2017}\cite{Aggarwal2012} with a related probabilistic Ansatz. In \cite{Beaujean2017}, the authors are assuming equal weights in their calculation and assume Jeffreys Prior to basically derive a distribution in $\lambda_i$. In \cite{Aggarwal2012}, the authors discuss a special case of the methodology described in section \ref{sec:poisson}, also using equal weights per bin and again assuming Jeffreys Prior. 

In the following section we will derive the natural probabilistic solution for $P(\lambda_i)$, and we will show that the involved Prior distribution depends on one free parameter, for which there exists a special "unique" value which we use throughout the paper. This value is shown to perform favorable against other alternatives, for example Jeffreys Prior. We then calculate the integral over $\lambda_i$ in closed form. Afterwards, the same principle idea is applied to the multinomial likelihood (section \ref{sec:multinomial}), where the analogous integration happens over the bin probabilities $p_i$ instead of bin expectation values $\lambda_i$.

\section{Finite-sample Poisson likelihood}
\label{sec:poisson}
\subsection{Equal weights per bin}
We start with the simpler case of equal weights for all Monte Carlo events.
We will also work with a single bin and drop the subscript $i$ without loss of generality. Given $k_{mc}$ events, each with weight $w$, one can perform Bayesian inference of the mean rate $\lambda$ given these Monte Carlo events via

\begin{align}
P(\lambda; k_{mc}, \alpha,\beta)&= \frac{ \mathrm{\textbf{P}}(k_{mc}; \lambda) \cdot \mathrm{\textbf{G}}(\lambda; \alpha, \beta)}{\int \mathrm{\textbf{P}}(k_{mc}; \lambda) \cdot \mathrm{\textbf{G}}(\lambda; \alpha, \beta) \ d\lambda} \label{eq:bayes} \\
&=\frac{e^{-\lambda(1+\beta)} \cdot \lambda^{k_{mc}+\alpha-1}}{(k_{mc}+\alpha-1)!} \cdot (1+\beta)^{k_{mc}+\alpha} \nonumber \\
&= \mathrm{\textbf{G}}(\lambda; k_{mc}+\alpha, 1+\beta) \equiv \mathrm{\textbf{G}}(\lambda; \alpha^{*},\beta^{*}) \nonumber
\end{align}
where $P(\lambda; k_{mc}, \alpha,\beta)$ is the Posterior, $ \mathrm{\textbf{P}}(k_{mc}; \lambda)$ the Poisson likelihood and $\mathrm{\textbf{G}}(\lambda; \alpha, \beta)$ a gamma distribution Prior with hyperparameters $\alpha$ and $\beta$. The use of the gamma Prior allows to write down closed-form expressions, while different choices of $\alpha$ and $\beta$ basically allow to model the whole spectrum of different Prior assumptions. The final expression has a closed-form solution and is again a gamma distribution with updated parameters $\alpha^{*}$ and $\beta^{*}$. Next, we require consistency conditions and knowledge about the final expression to fix $\alpha$ and $\beta$.
The inverse of the "rate" parameter $\beta$ in a gamma distribution acts as a scaling parameter for $\lambda$, exactly what the weight of each Monte Carlo event is doing already by definition. Therefore, we require the Posterior to scale with the Monte Carlo weight $w$, i.e. $1/w=\beta^*=1+\beta$, or $\beta=(1-w)/w$.
To pin  down $\alpha$, we could require that the final Posterior should have a mean value given by the number of Monte Carlo events weighted by their weight, i.e. $k_{mc} \cdot w$. The mean of the gamma Posterior distribution $\mathrm{\textbf{G}}(\lambda;\alpha^{*},\beta^{*})$ is analytically given by $\alpha^{*}/\beta^{*}=(k_{mc}+\alpha) \cdot w $, which corresponds to the desired value only if $\alpha=0$. Instead of forcing the mean of the Posterior to be $k_{mc} \cdot w$, we could have chosen to do the same procedure with the mode of $\mathrm{\textbf{G}}(\lambda;\alpha^{*},\beta^{*})$, which would result in $\alpha \neq 0$.
However, there is a second way to show that $\alpha=0$ is a somewhat special choice, which supports to take the mean in the preceding argumentation. To find it, we can imagine doing inference for all MC events individually, i.e. $k_{mc}=1$, and afterwards performing the convolution of the individual Posterior distributions. This is the generalization that has been hinted at already in eq. (\ref{eq:base_formula_with_convolution_of_gammas}). The result of this convolution must be the same as the original expression using all MC events. Using the known result that the convolution of two gamma distributions with similar rate parameter is $\mathrm{\textbf{G}}(\lambda; a_1,b) \ast \mathrm{\textbf{G}}(\lambda; a_2,b) = \mathrm{\textbf{G}}(\lambda; a_1+a2,b)$, we find

\begin{align}
P(\lambda)&=\mathrm{\textbf{G}}_1(\lambda; 1+\alpha^{\prime}, 1/w) \ast \ldots \ast \mathrm{\textbf{G}}_{k_{mc}}(\lambda; 1+\alpha^{\prime}, 1/w) \label{eq:convolution_equal} \\
&= \mathrm{\textbf{G}}(\lambda; k_{mc}+k_{mc}\cdot \alpha^{\prime}, 1/w) \nonumber \\
&\stackrel{!}{=} \mathrm{\textbf{G}}(\lambda; k_{mc}+ \alpha, 1/w)  \label{eq:convolution_prior_parameter}
\end{align}
using  $\alpha^{\prime}$ to denote that the Prior could in principle be different for the inference from individual events. We therefore only have consistency if $\alpha^{\prime}=\alpha/k_{mc}$. After some contemplation, this is reasonable: we encode the knowledge of the number of Monte Carlo events in the Prior for individual events. Nonetheless, one still has to determine $\alpha$. However, there now is one special choice, namely $\alpha^{\prime}=\alpha=0$, where we have compatibility between both scenarios, and exactly the same choice of Prior. This is the same value that has been derived from the previous scaling requirement of the mean. Also, there is no dependence on the overall number of Monte Carlo events in the Prior for individual events before the convolution. We call this choice of Prior "unique", and as was shown before, the corresponding Posterior is a gamma distribution whose mean is $w \cdot k_{mc}$. Let us quickly discuss the role of the gamma distribution, and why it seems to be the natural, and possibly only choice, to take the role in the generalization in eq. \ref{eq:base_formula_with_convolution_of_gammas}.
\begin{enumerate}
	\item The gamma distribution is the resulting Posterior for Bayesian inference with a Poisson likelihood for all major Prior assumptions, and simultaneously for Frequentist confidence intervals from the likelihood ratio (flat Prior). 
	\item The convolution of gamma distributions with the same scale corresponds again to a gamma distribution. This  property has to be fulfilled in order to be consistent again with Bayesian inference / Frequentist confidence intervals if multiple MC events are present with the same weight (see eq. (\ref{eq:convolution_equal}) and eq. (\ref{eq:convolution_prior_parameter})).
	\item Weighted normalized sums of gamma random variables correspond to Dirichlet-like random variables, and naturally extend the whole procedure to the multinomial setting (see section \ref{sec:multinomial}).
	\item The integral over the convolution of gamma distributions and a Poisson factor (eq. \ref{eq:base_formula_with_convolution_of_gammas}) involves hypergeometric functions, a new probability distribution, and is analytically tractable, even when the weights are different. This will be discussed in section \ref{sec:poisson_general_weights}.
\end{enumerate}

The Posterior $P(\lambda)$ (eq. \ref{eq:convolution_prior_parameter}) written out reads

\begin{align}
{P_{\mathrm{equal}}(\lambda; k_{mc}^{*}, w)}&=\mathrm{\textbf{G}}(\lambda; k_{mc}^{*}, 1/w) \\ &=\frac{e^{-\lambda \cdot (1/w)} \cdot \lambda^{k_{mc}^{*}-1}}{\Gamma(k_{mc}^{*})} \cdot (1/w)^{k_{mc}^{*}} \label{eq:plambda_equal}
\end{align}
with $k_{mc}^{*}=k_{mc}+\alpha$ to encode the Prior freedom for $\alpha$. We will use this definition of $k_{mc}^{*}$ for the rest of the paper. In general, $k_{mc}^{*}$ must be a positive real number for the definition of the gamma function, so $\alpha>-1$. For the unique Prior, $k_{mc}^{*}=k_{mc}$, i.e. $k_{mc}^{*}$ is a positive integer, and the gamma function $\Gamma(k_{mc}^{*})$ becomes a factorial.
It is important to mention that the choices $\alpha=0$ and $\beta=(1-w)/w$ for $w>1$ can not be used explicitly in Bayes' theorem (eq. \ref{eq:bayes}), since the gamma distribution requires $\alpha>0$, $\beta>0$. However, the problematic factors cancel out in the Posterior formation, and the end result is well defined. 

\begin{figure}
	\centering
	\includegraphics{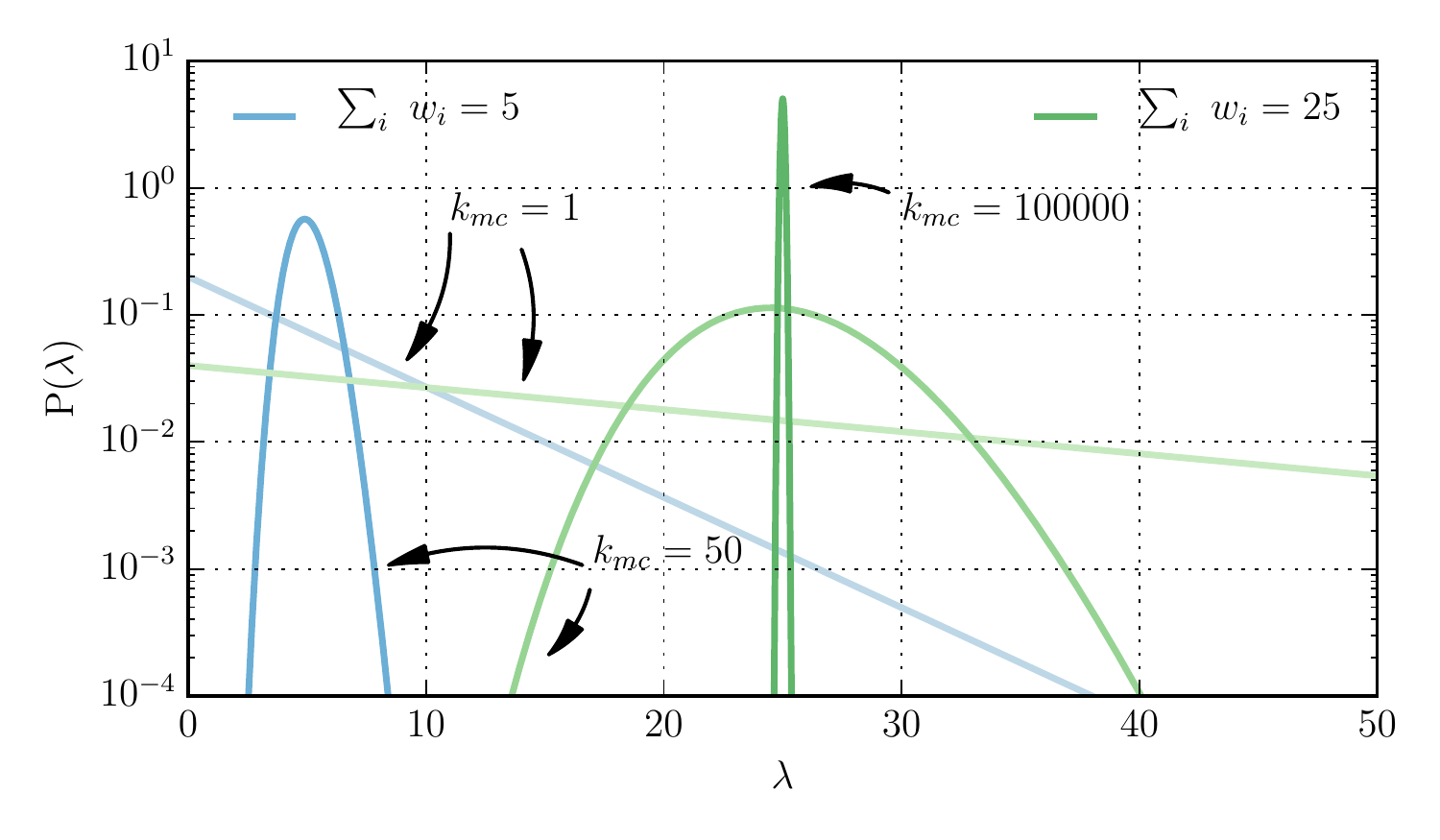}
	\caption{Examples of the distribution $P_{\mathrm{equal}}(\lambda; k_{mc}^{*}, w)$ for the expectation value $\lambda$ using equal weights (eq. \ref{eq:plambda_equal}). The unique Prior is used, which means $k_{mc}^{*}=k_{mc}$. The distribution for $\sum w_i=5$ (blue) is shown for $1$ and $50$ MC events. The distribution for $\sum w_i=25$ (green) is shown for $1$, $50$ and $100000$ MC events.}
	\label{fig:equal_behavior}
\end{figure}

With the Prior distribution fixed, we can now ask if the correct limiting behavior is achieved for $k_{mc} \rightarrow \infty$, $w \rightarrow 0$, which should be a delta distribution as shown in eq. (\ref{eq:poisson_to_expectation_value}). Indeed, for $k_{mc} \rightarrow \infty$, $w \rightarrow 0$ the mean asymptotically behaves as $\mathrm{E}[\lambda]_P = k_{mc}^{*}\cdot w \approx \sum_i w_i$ and the variance as $\mathrm{Var}[\lambda]_P = k_{mc}^{*} \cdot w^2 \approx (\sum w_i) \cdot w \rightarrow 0$, since the Prior freedom $\alpha$ can be neglected for large counts. Examples of $P(\lambda; k_{mc}^{*},w)$ are shown in figure \ref{fig:equal_behavior} for two different total sums of weights and different amounts of sample events $k_{mc}$ for the unique Prior.
As the number of sampling events $k_{mc}$ increases, one can see how the distribution approaches a delta peak centering around the sum of weights.
To obtain the modified likelihood expression, we use eq. (\ref{eq:poisson_to_expectation_value}), which is solved analytically using eq. (\ref{eq:exp_value_solution}) in the appendix. The overall calculation reads

\begin{align}
L_{\mathrm{\textbf{P}, finite, eq.}} &=
 \mathrm{E}\left[\frac{{e^{-\lambda}}{\lambda}^{k} }{k!}\right]_{{P_{\mathrm{equal}}(\lambda; k_{mc}^{*}, w)}} \nonumber \\
& = \frac{(1/w)^{k_{mc}^{*}}  \cdot \Gamma(k+k_{mc}^{*})}{  \Gamma(k_{mc}^{*}) \cdot k!\cdot(1+1/w)^{k+k_{mc}^{*}}} \label{eq:equal_form_one}  \\
 &\stackrel[\alpha=0]{}{=}  \frac{(\frac{k_{mc}}{\sum_j {w_j}})^{k_{mc}}  \cdot (k+k_{mc} - 1)!}{(k_{mc} - 1)! \cdot k!\cdot(1+\frac{k_{mc}}{\sum_j {w_{j}}})^{k+k_{mc}}}  \label{eq:equal_form_two} \\
 &\stackrel[\mathclap{ \substack{ k_{mc} \rightarrow \infty \\ \sum_j w_j \rightarrow const.}}]{}{=} \frac{e^{-\sum_j w_{j}} \cdot (\sum_j w_{j})^{k}}{k!} \nonumber
\end{align}
Since all weights are equal, $1/w=\frac{k_{mc}}{\sum_j w_j}$.  The expression for all bins is a multiplication of this factor for each bin, just as for the usual situation of independently distributed Poisson data. Equation \ref{eq:equal_form_one} can also be used as an approximate formula even if not all weights are equal, with $w$ being the mean weight of all weights in the bin. Equation \ref{eq:equal_form_two}, in this case shown for the unique Prior $\alpha=0$, is useful to derive the limiting standard Poisson behavior for $k_{mc} \rightarrow \infty$ and individual weights $w \rightarrow 0$. Other authors have considered eq. \ref{eq:equal_form_one} previously in this context for the special case of Jeffreys Prior ($\alpha=0.5$) in \cite{Aggarwal2012}.

\subsection{General weights}
\label{sec:poisson_general_weights}

\begin{figure}
	\centering
	\includegraphics{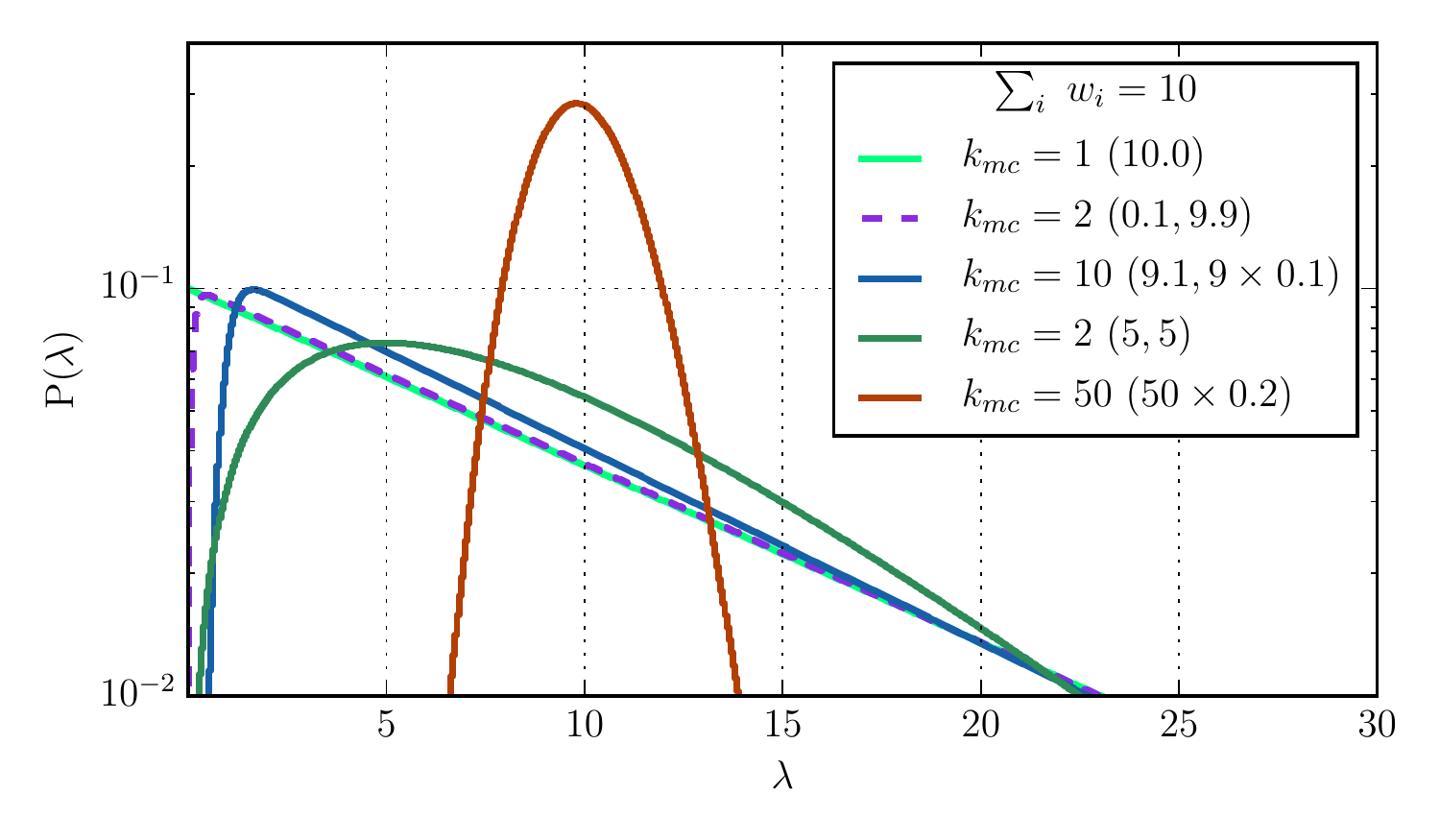}
	\caption{Examples of the distribution $P_{\mathrm{general}}(\lambda; \bm{k_{mc}^{*}}, \bm{w})$ for the expectation value $\lambda$ using arbitrarily weighted Monte Carlo events (eq. \ref{eq:p_lambda_general}). The unique Prior is used, which means $k_{mc,i}^{*}=k_{mc,i}$. The values in parenthesis indicate the weight distribution of these events, while the sum of all weights is always equal to $10$.}
	\label{fig:general_behavior}
\end{figure}

The generalization to different weights per MC event follows directly from eq. (\ref{eq:convolution_equal}) using arbitrary and possibly different weights for each factor in the convolution.
Convolutions of general gamma distributions arise in many applications, e.g. in composite arrival time data \cite{Sim1992} or composite samples with weighted events \cite{DiSalvo98}. The general solution can not be written down in a closed-form expression, but one can for example write it in terms of a generalized confluent hypergeometric function \cite{DiSalvo2008} or via a perturbative sum \cite{Moschopoulos1985}. We choose the perturbation representation first (loosely following the notation in \cite{Moschopoulos1985}), because it allows to easily calculate a result to a given desired precision. In the following, we treat the slightly more general case where we assume there are N distinct weights among the MC events which are enumerated with the index $j$ and come with multiplicities $k_{mc,j}$, the total number of MC events being $k_{mc}$. Then, the final expression for the convolution of $N$ general gamma-distributed Posteriors reads

\begin{align}
P_{\mathrm{general}}(\lambda; \bm{k_{mc}}, \bm{w})=C \cdot \sum_{l=0}^{\infty} {\delta_l \cdot \frac{ \lambda^{\rho+l-1} \cdot e^{-\lambda/w_N  }   }{ (\rho+l-1)! \cdot w_N^{\rho+l} } }   \label{eq:p_lambda_general}
\end{align},
where $w_N$ is the smallest weight in the bin, $\rho=\sum_j k_{{mc},j}^{*} = k_{{mc}}+\alpha$, $C=\prod_{j=1}^{N} (\frac{w_N}{w_j})^{k_{{mc},j}^{*}}$, $\gamma_k=\sum_{j=1}^{N} k_{mc,j}^{*} \frac{(1-w_N/w_j)^k}{k} $ and $\delta_{k+1}=\frac{1}{k+1} \sum_{i=1}^{k+1}{i \cdot \gamma_i \cdot \delta_{k+1-i}}$ for positive integer k which is constructed iteratively with $\delta_0=1$ to the desired order. The term $k_{{mc},j}^{*}=k_{{mc},j}+k_{{mc},j}\cdot \alpha/k_{mc}$ corresponds to the definition in eq. (\ref{eq:convolution_equal}), but generalized to $k_{{mc},j}$ weights per individual factor in the convolution.  
The overall behavior can be seen in figure \ref{fig:general_behavior}, showing $P_{\mathrm{general}}(\lambda)$ for different examples of MC events which always sum to $10$. 
The distribution is usually dominated by events that are close to each other, as can be seen by the example with nine small weights and one large one (blue curve). The expression  reduces to equation (\ref{eq:plambda_equal}) when all weights are equal, since then $\delta_0=1$ and $\delta_l=0 \  \forall \ l>0$. In the other extreme, as the spread between the weights gets larger, more and more terms in the perturbation series have to be taken into account in order to come close to the exact result. Since $\delta$, and thereby $P_{\mathrm{general}}(\lambda)$, is constructed iteratively, we can define a convenient stopping criterion for the iterative calculation. The desired relative precision $p$ can be reached by truncating the series when $C \cdot \sum_{i} \delta_{i} \approx p$, since $C \cdot \sum_{i = 0}^{\infty} \delta_{i}=1$. 

The next step is to form the corresponding generalization of the marginal likelihood for equal weights (eq. \ref{eq:equal_form_one}), i.e. integrating a Poisson factor mean $\lambda$ with $P_{\mathrm{general}}(\lambda)$. One possibility is to pull the integration inside the sum from eq. (\ref{eq:p_lambda_general}), and continue analogously to the equal-weights case. The result is an infinite series-representation of the marginal likelihood, and its calculation is shown in appendix \ref{appendix:series_derivation}.  Here, however, we want to continue with the previously mentioned representation of $P_{\mathrm{general}}(\lambda)$ in terms of a generalized confluent hypergeometric function, since it is possible to derive an analytic closed-form expression along those lines. In this representation, $P_{\mathrm{general}}(\lambda)$ looks like \cite{DiSalvo2008}

\begin{align}
\begin{aligned}
P_{\mathrm{general}}(\lambda) = \frac{\lambda^{k_{mc}^{*}-1} \cdot (1/w_N)^{k_{mc}^{*}} e^{-\lambda \cdot (1/w_N)}}{\Gamma(k_{mc}^{*})} \cdot \left( \prod_{j=1}^{N} \left(\frac{w_N}{w_j}\right)^{k_{mc,j}^{*}} \right) \\ \cdot \Phi_2^*(\bm{k_{mc, 1 \ldots N-1}^{*}};k_{mc}^{*};\lambda \cdot \bm{z^{*}} )
\end{aligned}
\end{align},
where $N$ is the number of distinct weights,  $w_N$ is the smallest weight, and $j$ enumerates like $j=1,\ldots,N-1$. The function $\Phi_2^*$ denotes the multidimensional generalization of the confluent Humbert series $\Phi_2$ with  ${z^*}_j=1/w_N - 1/w_j$. The corresponding Poisson marginal likelihood for one bin is again formed via the expectation value

\begin{align}
&\begin{aligned}
L_{\mathrm{\textbf{P},finite,gen.}} &=
\mathrm{E}\left[\frac{{e^{-\lambda}}{\lambda}^{k} }{k!}\right]_{P_{\mathrm{general}}(\lambda)} \nonumber  
\end{aligned}\\
&\begin{aligned}
&= \int_0^{\infty} \frac{\lambda^{k_{mc}^{*}+k-1} \cdot (1/w_N)^{k_{mc}^{*}} \cdot e^{-\lambda \cdot (1 + 1/w_N)}}{\Gamma(k_{mc}^{*}) \cdot k!}  \\ & \cdot \left( \prod_{j=1}^{N} \left(\frac{w_N}{w_j}\right)^{k_{mc,j}^{*}} \right) \cdot \Phi_2^*(\bm{k_{mc, 1 \ldots N-1}^{*}};k_{mc}^{*};\lambda \cdot \bm{z^*} ) \ d\lambda 
\end{aligned}\\
&\begin{aligned}
&=\frac{\Gamma(k_{mc}^{*}+k)}{\Gamma(k_{mc}^{*}) \cdot k!} \cdot \left(\frac{1}{w_N}\right)^{k_{mc}^{*}} \cdot \left(\frac{1}{1+1/w_N}\right)^{k_{mc}^{*}+k} \cdot \left( \prod_{j=1}^{N} \left(\frac{w_N}{w_j}\right)^{k_{mc,j}^{*}} \right) \\ &\cdot  \int_0^{\infty} {\lambda^*}^{k_{mc}^{*}+k-1}  e^{-\lambda^*} \cdot  \Phi_2^*(\bm{k_{mc, 1 \ldots N-1}^{*}};k_{mc}^{*};\frac{\lambda^*}{1+1/w_N} \cdot \bm{z^*} ) \ d\lambda^* \label{eq:laplace_type} 
\end{aligned} 
\\
&\begin{aligned}
&=  \frac{\Gamma(k_{mc}^{*}+k)}{\Gamma(k_{mc}^{*}) \cdot k!} \cdot \left(\frac{1}{w_N}\right)^{k_{mc}^{*}} \cdot \left(\frac{1}{1+1/w_N}\right)^{k_{mc}^{*}+k} \\ &\cdot \left( \prod_{j=1}^{N} \left(\frac{w_N}{w_j}\right)^{k_{mc,j}^{*}} \right) \cdot F_D(k_{mc}^{*}+k;\bm{k_{mc, 1 \ldots N-1}^{*}};k_{mc}^{*};\bm{z^{**}}) 
\end{aligned}
\label{eq:laplace_type_lauricella_replaced}
\end{align}
with $z^{**}_j = 1-\frac{1+1/w_j}{1+1/w_N}$. A variable transform $\lambda^*=\lambda \cdot (1+1/w_N)$ is used to form a Laplace-type integral (eq. \ref{eq:laplace_type}) which is equivalent to the fourth Lauricella function $F_D$ via $F_D(a;\bm{b};c;\bm{z})=\int_0^{\infty} t^{a-1} e^{t}  {\Phi_2^*}(\bm{b};c;t \cdot \bm{z}) \ dt$ \cite{1976Exton}. The fourth Lauricella function is a certain extended form of the Gauss hypergeometric function, and appears for example in statistical problems \cite{Dickey1983} or even theoretical physics \cite{1972Miller}. For $F_D(a;\bm{b};c;\bm{z})$ with $\sum_i b_i <c$, as it is the case here, it is possible to evaluate $F_D$ via numerical integration of an integral representation \footnote{See for example equation $1.9$ in \cite{1974Hattori}} over the simplex. However, this is not much more practical than the previously discussed series representation derived in appendix \ref{appendix:series_derivation} - both require long evaluation times for larger number of weights if the end result is supposed to be accurate, especially if the relative variation in the weights is large. However, for $a-c$ being a non-negative integer, which is the also the situation here, we can rewrite a generic $F_D$ via

\begin{align}
F_D(a;\bm{b};c;\bm{z}) &=R_{-a}(\bm{b_{+1}}, 1-\bm{z_{+1}}) \nonumber \\
&=\left( \prod_i (1-z_{+1,i})^{-b_{+1,i}} \right) \cdot R_{a-c} \left(\bm{b}_{+1}, (1-\bm{z}_{+1})^{-1}\right) \nonumber \\
&\begin{aligned}
&=\left(\prod_i (1-z_{+1,1})^{-b_{+1,i}} \right) \\ &\cdot \left(\stackrel[\sum_i k_i=a-c, \ k_i \geq 0]{}{\sum} \mathrm{\textbf{DM}}(\bm{k}; \bm{b_{+1}}) \prod_i (1-z_{+1,1})^{-k_{i}} \right) \label{eq:multinomial_rep_for_fd}
\end{aligned} 
\end{align}
 , where $F_D$ is expressed via Carlson's Dirichlet average $R_n$ \cite{Carlson1963} in the first step. In the second step, we modify its first parameter \cite{Carlson1977} \cite{Neuman1994a} from $-a$ to $a-c$, which picks up an extra factor. In the switch from $F_D$ to $R_n$, one always adds one element to $\bm{b}$ and $\bm{z}$, i.e. $\bm{z_{+1}}=[0, \bm{z}]$ and $\bm{b_{+1}}=[b_0, \bm{b}]$ with $b_0=c-\sum_i b_i$, reflecting the homogeneity properties of $R_n$ \cite{Carlson1963}. Looking back at eq. (\ref{eq:laplace_type_lauricella_replaced}), we can identify $b_0$ with $k_{mc,N}^{*}$, i.e. the  multiplicity and Prior factor of the smallest weight, which has been left out in $\bm{k_{mc, 1 \ldots N-1}^{*}}$ in the Lauricella function before, and subsequently $\bm{b_{+1}}=\bm{k_{mc}^{*}}$. 
In the last step we exploit that $R_n$ with non-negative integer $n$ is the probability generating function of a Dirichlet-multinomial distribution \cite{Dickey1983}, which applies since $x=a-c=k_{mc}^{*}+k-k_{mc}^{*}=k$ is a non-negative integer. The Dirichlet-multinomial distribution is denoted by the factor $\mathrm{\textbf{DM}}$. It can be derived as the expectation value of a multinomial distribution under a Dirichlet distribution \cite{Balakrishnan1997}, and looks like
 
 \begin{align}
\mathrm{\textbf{DM}}(\bm{k}; \bm{k_{mc}^{*}})=\frac{k! \cdot \Gamma(k_{mc}^{*})}{\Gamma(k+k_{mc}^{*})}  \prod_{i} \frac{\Gamma(k_i+k_{mc,i}^{*})}{k_{i}!\cdot \Gamma(k_{mc,i}^{*})} \label{eq:multinomial_dirichlet1}
 \end{align} with $k=\sum_i k_i$ and $k_{mc}^{*}=\sum_i k_{mc,i}^{*}$. It will appear again later in the context of the generalization of the multinomial likelihood, where it might be easier to put it into context.

 We will now simplify eq. (\ref{eq:multinomial_rep_for_fd}) further. First we observe that parts of the combinatorial sum can be written in complex analysis form \cite{Zhou2011}, namely
 
 \begin{align}
 \stackrel[\sum_i k_i=K, \ k_i \geq 0]{}{\sum}{ \prod_i^{N} x_i^{k_i}} = \frac{1}{2 \pi i} \oint \frac{t^{N+K-1}}{\prod_i (t-x_i)}  dt \label{eq:combinatorial_sum_complex}
 \end{align}
 where the contour integral is over the circle containing all $x_i$ on the inside. Next, we write $F_D$ for the case of the unique Prior and assume that all weights are different, i.e. $b_{+1,i}=k_{mc,i}^{*}=k_{mc,i}=1$.  We write this as $\bm{b_{+1}}=\bm{1}=[1,\ldots,1]$), and for convenience use a modified argument \footnote{These are the indivdiual components of the vector argument, which is written as $1-\bm{z}^{-1}=[1-z_1^{-1}, 1-z_2^{-1}, \ldots]$} $z_i \rightarrow 1-z_i^{-1}$, which results in
 
 \begin{align}
 \begin{aligned}
F_D(a;\bm{1}_{1 \ldots N-1};c;1-\bm{z}^{-1}) &= F_D(k+k_{mc};\bm{1}_{1 \ldots N-1};k_{mc};1-\bm{z}^{-1}) \nonumber
\end{aligned}  \\
\begin{aligned}
&= \left(\prod_i^{k_{mc}} (z_{+1,i})^{1} \right) \cdot \left(\stackrel[|\bm{k}|=k, \ k_i \geq 0]{}{\sum} \mathrm{\textbf{DM}}(\bm{k}; \bm{1}) \prod_i^{k_{mc}} (z_{+1,i})^{k_{i}} \right) \nonumber
\end{aligned} \\
\begin{aligned}
&=\mathrm{\textbf{DM}}(\bm{k}; \bm{1}) \cdot \left(\prod_i^{k_{mc}} z_{+1,i} \right) \cdot \frac{1}{2 \pi i} \oint \frac{t^{k_{mc}+k-1}}{\prod_i^{k_{mc}} (t-z_{+1,i})}  dt \label{eq:fd_as_contour}
\end{aligned} \\
\begin{aligned}
&=\frac{\Gamma(a-c+1) \cdot \Gamma(c)}{\Gamma(a)} \cdot \left(\prod_i^{c} z_{+1,i} \right) \cdot \frac{1}{2 \pi i} \oint \frac{t^{a-1}}{\prod_i^{c} (t-z_{+1,i})}  dt \nonumber
\end{aligned}
\end{align}
, where we first use eq. (\ref{eq:multinomial_rep_for_fd}) to rewrite $F_D$ as a combinatorial sum. In the second step the Dirichlet-multinomial factor $\mathrm{\textbf{DM}}(\bm{k}; \bm{1})$ is pulled out of the combinatorial sum in since it is constant for $b_{+1}=k_{mc,i}^{*}<2$, and we use eq. (\ref{eq:combinatorial_sum_complex}) to rewrite the whole expression via a contour integral. Just as a reminder, $a=k+k_{mc}$ and $c=k_{mc}$, as defined earlier. 
 All weights $w_i$ are distinct by construction, so the $z_i$ are distinct as well since $z_i=\frac{1+1/w_N}{1+1/w_i}$ in the new nomenclature of $z_i$. We can now imagine two weights $w_i$, and thereby also the corresponding $z_i$, approach each other more and more, until they merge into a double pole in the contour integral (eq. \ref{eq:fd_as_contour}) and the prefactor becomes $z_{+1,i} \rightarrow z_{+1,i}^2 $. In general for $k_{mc,i}$ weights with equal weight, the pole becomes a pole with multiplicity $k_{mc,i}$ and the general expression for $F_D$ with $a>c$ and $c=\sum b_{+1,i}$ behaves as
  
 \begin{align}
 &F_D(a;\bm{b};c;1-\bm{z}^{-1}) \nonumber \\
 &= \frac{\Gamma(a-c+1) \cdot \Gamma(c)}{\Gamma(a)}  \cdot \left(\prod_i^{N} {z_{+1,i}}^{b_{+1,i}} \right) \cdot \frac{1}{2 \pi i} \oint\limits_{\rho=R\rightarrow \infty} \frac{t^{a-1}}{\prod_i^{N} (t-z_{+1,i})^{b_{+1,i}}}  dt \label{eq:fd_as_residue} \\
 &= \frac{\Gamma(a-c+1)  \cdot \Gamma(c)}{\Gamma(a)}  \cdot (-1)^{b_{+1,i}} \cdot \frac{1}{2 \pi i}  \cdot  \oint\limits_{\rho=\epsilon} \frac{1}{t^{a-c+1} \cdot \prod_i^{N} (t-1/{z_{+1,i}})^{b_{+1,i}}}  dt \label{eq:fd_as_residue_infinity} \\
 &= \frac{\Gamma(a-c+1)  \cdot \Gamma(c)}{\Gamma(a)} \cdot \left(\prod\limits_{i=1}^{N} {z_{+1,i}}^{b_{+1,i}}\right) \cdot D_{a-c}(\bm{b_{+1}}, \bm{z_{+1}}) \label{eq:mathematical_newdef_lauricella}
 \end{align}
, where $N =\mathrm{len}(\bm{b})+1=\mathrm{len}(\bm{b_{+1}})$,  $z_{+1,N}=1$, $b_{+1,N}=c-\sum_i b_i$ and the factor $D_j$ is iteratively defined via $D_{j}= \frac{1}{j}\sum\limits_{k=1}^{j} \left[\left(\sum\limits_{i=1}^{N} b_{+1,i} \cdot {z_{+1,i}}^k \right) D_{j-k}\right]$ with $D_0=1$. We first change the contour integral to a form which calculates the residue at infinity (eq. \ref{eq:fd_as_residue_infinity}), which we subsequently write as a finite sum $D_{j}$ using the algorithm in \cite{Ma2014}. Because we evaluate the integral with the residue at infinity, we can again allow for a Prior $\alpha \neq 0$, which generalizes $k_{mc} \rightarrow k_{mc}+\alpha = k_{mc}^{*}$. Thus, the above form of $F_D$ can be extended again to real $a$, $\bm{b}$ and $c$ (or to $k_{mc}^{*}$, $\bm{k_{mc}}^{*}$ and $k_{mc}^{*}+k$ in the other nomenclature for this concrete problem), retaining only the constraint that $a-c$ must be a non-negative integer. 

To summarize, the reformulations of $F_D$ allow to write the single-bin Poisson likelihood incorporating uncertainty from finite Monte Carlo data (eq. \ref{eq:laplace_type_lauricella_replaced}) as
 
\begin{align}
L_{\mathrm{\textbf{P},finite,gen.,1}} =  \stackrel[|\bm{k}|=k, \ k_i \geq 0]{}{\sum}  \prod_i \frac{ \Gamma(k_i+k_{mc,i}^{*})  }{k_i!\cdot \Gamma(k_{mc,i}^{*})} \cdot \left(\frac{1}{w_i}\right)^{k_{mc,i}^{*}} \cdot \left(\frac{1}{1+1/w_i} \right)^{k_i+k_{mc,i}^{*}} \label{eq:finite_poisson_combinatorial}
\end{align}
or

\begin{align}
L_{\mathrm{\textbf{P},finite,gen.,2}} =  \left(\prod_i \left(\frac{1}{1+w_i}\right)^{k_{mc,i}^*} \right) \cdot D_k(\bm{k_{mc}^*},\frac{1}{1+1/\bm{w}}) \label{eq:finite_poisson_finitesum}
\end{align}, by plugging in eq. (\ref{eq:multinomial_rep_for_fd}) or eq. (\ref{eq:mathematical_newdef_lauricella}),  respectively. The term $D_k$ is again iteratively defined as in eq. (\ref{eq:mathematical_newdef_lauricella}). The index $i$ goes over all distinct weights. It is quite remarkable, that in the latter representation all gamma factors cancel out, and one is left with one weight prefactor and an iterative finite sum $D_k$. For multiple bins, the full likelihood can be constructed as the product of the likelihood for each individual bins, similar to the standard Poisson likelihood for independently distributed data.
The combinatorial sum (eq. \ref{eq:finite_poisson_combinatorial}) has an intuitive interpretation as marginalization of permutations, running over all possible combinations of counts $k_i$ distributed among the weights, and for each weight we have a factor corresponding to the  result we earlier derived for a single weight (eq. \ref{eq:equal_form_one}).
While this combinatorial expression very quickly becomes unmanageably large, the finite sum (eq. \ref{eq:finite_poisson_finitesum}) has a substantial computational advantage and is more usable in practice. Due to the relation to Carlson's Dirichlet average $R_n$ in eq. (\ref{eq:multinomial_rep_for_fd}), the end result is also an efficient way to calculate several mathematical quantities including the probability generating function (PGF) of the Dirichlet-multionomial distribution, the general divided difference of a monomial function \cite{book:divdiv} or moments of univariate B-Splines \cite{Carlson1991}. More information can be found in appendix \ref{appendix:pgf_and_others}.

\section{Finite-sample multinomial likelihood}
\label{sec:multinomial}
Let us calculate the analogous finite-sample expression of the multinomial likelihood. The compound distribution in the multinomial case comes from the integration of bin probabilities $p_i$ instead of expectation values $\lambda$,

\begin{align}
L_\mathrm{\textbf{MN},finite}&= \mathrm{E}\left[{\textbf{MN}}(k_1, \ldots, k_{N}; p_1, \ldots p_N)\right]_{P(p_1, \ldots, p_N)} \\ &= \stackrel[\sum p_i=1]{}{\int_{p_1} \ldots \int_{p_{N}}} {\textbf{MN}}(k_1, \ldots, k_{N}; p_1, \ldots p_N) \cdot P(p_1 \ldots, p_N) \ dp_1 \ldots dp_{N} \label{eq:simplex_full} \\
&\begin{aligned}
=\frac{k!}{k_1! \ldots k_{N}!}  \stackrel[\sum p_i\leq1]{}{\int_{p_1} \ldots \int_{p_{N-1}}} p_1^{k_1} \cdot \ldots \cdot {p_{N-1}}^{k_{N-1}} \cdot \left(1-\sum_i^{N-1} {p_i}\right)^{k_N} \\ \cdot P(p_1, \ldots, p_{N-1}) \ dp_1 \ldots dp_{N-1} \label{eq:simplex_m1}
\end{aligned}
\end{align}
with $N$ is equal to the number of bins. The integration happens over the $N-1$ simplex, and either has the constraint $p_1 + \ldots p_N = 1$ in eq. (\ref{eq:simplex_full}) or $p_1+ \ldots+ p_{N-1} \leq 1$ in eq. (\ref{eq:simplex_m1}). The latter one is easier to work with in practice and we will usually do so in the rest of the paper. 

\subsection{Equal weights per bin}
First, let us look at equal weights per bin. The analogue to the gamma distribution in the Poisson situation corresponds here to the scaled Dirichlet distribution \cite{Monti2011}

\begin{align}
\begin{aligned}
P(p_1, \ldots, p_{N-1})&= \frac{\Gamma(\alpha_{tot}^{*})}{\prod_i^{N} \Gamma(\alpha_i^{*}) }  \cdot  \left( \prod_i^{N} {\beta_i^{*}}^{\alpha_i^{*}} \right) \cdot \frac{ {p_1}^{\alpha_1^{*}-1} \ldots {p_{N-1}}^{\alpha_{N-1}^{*}-1} }{ \left(\beta_N^{*} (1-\sum_i^{N-1} p_i) + \sum_i^{N-1} \beta_i^{*} \cdot p_i\right)^{\alpha}} \\ &\cdot \left(1-\sum_i^{N-1} {p_i}\right)^{\alpha_N^{*}-1} 
\end{aligned}
\end{align}, which can be derived as gamma random variables (one for each bin) which are each normalized according to their sum. To be consistent with the Poisson case, we again have $\alpha_i^{*}=k_{mc}^{*}=k_{mc,i}+\alpha_i$ and  $\beta_i^{*}=1/w_i$ parameters that have a similar meaning as before, with the difference that $i$ now always stands for individual bins. Since the Prior should not depend on the bin, we set $\alpha_i=\alpha$. For the multinomial likelihood, there is no single-bin viewpoint, except the trivial one. For the parameter $\alpha_{tot}^{*}=\sum_i \alpha_i^{*}=k_{mc} + N \cdot \alpha$, we are in a similar consistency dilemma as before with the Poisson derivation - its value depends on the number of bins $N$, and would increase as more and more bins are taken into account.  This means if we did not want this value to change with increased number of bins, we would have to go the other way around by defining $\alpha_{tot}^{*}=k_{mc}+\alpha$ and then redefine $\alpha_i^{*}=k_{mc,i}+\alpha/N$. Here, if more bins were used, it would be taken into account in the $\alpha_i^{*}$, but the overall $\alpha_{tot}^{*}$ would not change. Again, this dilemma can be solved by the unique Prior $\alpha=0$, where no such ambiguity exists, neither if more Monte Carlo events nor if more bins are being used. This issue will later become especially apparent for ratio-constructions (see section \ref{sec:ratio_constructions}). For now, we use the first definition to be comparable to the Poisson case.

When the weights in all bins are equal, i.e. all $w_i=w$, $P$ reduces to the standard Dirichlet distribution and the compound likelihood (solution to the integral in eq. \ref{eq:simplex_m1}) becomes the already in eq. (\ref{eq:multinomial_dirichlet1}) introduced Dirichlet-multinomial distribution $\mathrm{\textbf{DM}}$.

\begin{align}
L_\mathrm{\textbf{MN},finite,all \ equal}&=\mathrm{\textbf{DM}}(k_1,\ldots,k_N;k_{mc,1}^{*},\ldots, k_{mc,N}^{*}) \nonumber
\end{align}
The integration procedure using the general scaled Dirichlet density with different $\beta^{*}_i$ is more involved and calculated in detail in appendix \ref{appendix:marginal_llh_multinomial}. The final result looks like

\begin{align}
&\begin{aligned}
\mathllap{L_\mathrm{\textbf{MN},finite, eq.}} &=\stackrel[\sum p_i\leq1]{}{\int_{p_1} \ldots \int_{p_{N-1}}} \mathrm{\textbf{MN}}(k_1, \ldots, k_{N}; p_1, \ldots p_N) \\ \cdot \frac{\Gamma(\alpha_{tot}^{*})}{\prod_i^{N} \Gamma(\alpha_i^{*}) } & \cdot  \left( \prod_i^{N} {\beta_i^{*}}^{\alpha_i^{*}} \right) \cdot \frac{ {p_1}^{\alpha_1^{*}-1} \ldots {p_{N-1}}^{\alpha_{N-1}^{*}-1} }{ \left(\beta_N^{*} (1-\sum_i^{N-1} p_i) + \sum_i^{N-1} \beta_i^{*} \cdot p_i\right)^{\alpha_{tot}^{*}}} \\ & \cdot \left(1-\sum_i^{N-1} {p_i}\right)^{\alpha_N^{*}-1} \ dp_1 \ldots dp_{N-1} \label{eq:mnomial_scaled_dirichlet}
\end{aligned}\\
&\begin{aligned}
&= \mathrm{\textbf{DM}}(\bm{k}, \bm{k_{mc}^{*}}) \cdot \left( \prod_i^{N} {\left(\frac{w_N}{w_i}\right)}^{k_{mc,i}} \right) \cdot F_D(a;\bm{b};c,\bm{z}) \label{eq:finite_multinomial_equal}
\end{aligned}
\end{align}
, where $a=k_{mc}^{*}$, $b_i=k_{mc,i}^{*}+k_i \ (i=1 \ldots N-1)$, $c=k_{mc}^{*}+k$, $z_i=1-w_N/w_i \ (i=1 \ldots N-1)$.
The resulting probability distribution consists of a Dirichlet-multinomial factor, a factor consisting of the weights, and again the fourth Lauricella function $F_D$. Compared to the Lauricella function appearing the Poisson case, however, this $F_D$ has the first and third argument switched, and the second vectorial argument is generally larger. When all weights are equal, the expression is again a standard Dirichlet-multinomial distribution, since $F_D(a,\bm{b},c, \bm{0}) = 1$. We can not simplify $F_D$ similarly to the Poisson case, since $c=k_{mc}^{*}+k > a=k_{mc}^{*}$. For this situation, other specific finite-sum representations have been found by \cite{Tan2005}. However, they are a little more complicated in nature, and we do not write them out explicitly here.

For illustrative purposes, it is interesting to discuss that the Dirichlet-multinomial distribution $\mathrm{\textbf{DM}}(\bm{k}, \bm{k_{mc}^{*}})$ for integer parameters, in this case this means for the unique Prior $k_{mc}^{*}=k_{mc}$, corresponds to a so-called standard Polya-Urn model \cite{Johnson1977}. In this mental model, one draws differently colored balls from an urn, where the initial number of colored balls is fixed by a given color parameter. For each color drawn, one places the ball back, including an additional one of the same color. If one draws $K$ balls in this way, the $K$ balls are distributed according to the Dirichlet-multinomial distribution. This means, with the unique Prior, handling Monte Carlo simulations with equally weighted events in a multinomial evaluation with $N$ bins is exactly equivalent to a standard Polya-Urn modeling process with $N$ colors. In the multinomial Monte Carlo setup with equal weights per bin, the colors correspond to different bins, while the number of colored balls corresponds to the numbers of weighted Monte Carlo events $k_{mc,i}$ in a given bin $i$. 

\subsection{General weights}

For general weights, it is not clear what the analogue for $P(\lambda)$ would be. However, we can draw inspiration from the combinatorial expression for the finite-sample Poisson likelihood (eq. \ref{eq:finite_poisson_combinatorial}). We can imagine that every weight individually corresponds to a single imaginary bin. For the numerator, we then have to find all combinations of $k_{i, j}$, the counts in the imaginary bins $j$, such that $\sum_j k_{i, j}=k_i$,  where $k_i$ is the number of observed events in the real bin $i$. In the denominator, one relaxes this condition and sums over many more combinations with the only constraint that the total event count $k$ equals to the individual counts $k_{i,j}$ in the imaginary bins, independent of the real bin counts. The end result by construction has to be a probability distribution in $k_i$. The proposal distribution for the unique Prior that fulfills these criteria looks like

\begin{align}
\begin{aligned}
L_{\mathrm{\textbf{MN},finite, gen.}} =  \stackrel[\substack{\sum_i k_{1,i} = k_1  \\ \dots \\ \sum_i k_{N,i} = k_N \\ \sum_i k_i = k, \ k_i\geq 0}]{}{\sum} \mathrm{\textbf{DM}}(\bm{k}, \bm{1}) \cdot \left( \prod_i^{N} {\left(\frac{w_N}{w_i}\right)}^{1} \right) \\
  \cdot F_D(k_{mc};1+k_{1,1}, .,1+k_{1,M_1}, ., 1+k_{N-1,1} , ., 1+k_{N-1,M_{N-1}} ;k+k_{mc},\bm{z}) \\ \Bigg/ \stackrel[\substack{\sum_{i,j} k_{i,j} = k \\ k_{i,j} \geq 0  }]{}{\sum} \mathrm{\textbf{DM}}(\bm{k}, \bm{1}) \cdot \left( \prod_i^{N} {\left(\frac{w_N}{w_i}\right)}^{1} \right) \\
  \cdot F_D(k_{mc};1+k_{1,1}, .,1+k_{1,M_1}, ., 1+k_{N-1,1} , ., 1+k_{N-1,M_{N-1}} ;k+k_{mc},\bm{z})
 \end{aligned} \\
 \begin{aligned}
 = \frac{\stackrel[\substack{\sum_i k_{1,i} = k_1  \\ \dots \\ \sum_i k_{N,i} = k_N \\ \sum_i k_i = k, \ k_i\geq 0}]{}{\sum}  F_D(k_{mc};1+k_{1,1}, ., 1+k_{N-1,M_{N-1}} ;k+k_{mc},\bm{z}) }{\stackrel[\substack{\sum_{i,j} k_{i,j} = k \\ k_{i,j} \geq 0  }]{}{\sum}
 	 F_D(k_{mc};1+k_{1,1},., 1+k_{N-1,M_{N-1}} ;k+k_{mc},\bm{z})} \label{eq:finite_mnomial_general} 
 \end{aligned}
\end{align}
where $z_j=1-w_{\mathrm{N}}/w_j \ (j=1 \ldots N-1)$ and $w_\mathrm{N}$ is the smallest weight of all weights. The weight prefactor and Dirichlet-multinomial factor cancel out because every weight filled into the imaginary bins has single multiplicity $k_{mc,i}=1$ by construction for the unique Prior, i.e. the Dirichlet-multinomial prefactor is just a constant. This formulation is also motivated by the ratio representation discussed in the next section. It has been checked to be a proper probability distribution for non-trivial binomial-like problems numerically, but it is just a motivated construction, not a solid mathematical derivation as a generalization of the case for equal weights per bin. In practice, it is also not very useful because of the huge combinatorial calculation. It would be interesting to know if a simpler representation exists.

Finally, all results from the multinomial likelihood automatically carry over to an approximated unbinned likelihood similarly to eq. (\ref{eq:mnomial_unbinned_relation}) via a simple constant factor that depends on the binning. This can be useful in unbinned likelihood fits that combine analytic PDFs with MC-derived PDFs that are intrinsically binned and renormalized according to eq. (\ref{eq:mnomial_unbinned_relation}). However, one can see that a difference to the standard multinomial formula is only seen if the weights in the different bins are not all the same. Otherwise the multinomial likelihood corresponds to the Dirichlet-Multinomial distribution, which in the categorical case, the case which matters for the unbinned formula discussed in eq. (\ref{eq:mnomial_unbinned_relation}), is numerically equal to the standard multinomial formula for infinite statistics. Therefore, a difference can only be expected for MC-derived PDFs containing weighted simulation, not for data-derived PDFs where each event by definition has the same weight.

\section{Further finite-sample constructions}

With the generalized finite-sample expressions for the Poisson and multinomial case at hand, we can try to find relationships between the two. Let us recall that
the multinomial likelihood is similar to a division of Poisson factors with a single global Poisson factor

\begin{align}
L_{\mathrm{\textbf{MN}}}&=  \frac{\prod_{\mathrm{bins} \ i} L_{\mathrm{\mathbf{P}},i}}{L_{\mathrm{\mathbf{P},global}}}  
= \frac{\prod_{\mathrm{bins} \ i} \frac{e^{\lambda_i} \cdot \lambda^{k_i} }{{k_i}!}   }{\frac{e^{\lambda} \cdot \lambda^{k} }{{k}!}} \label{eq:poisson_mn_equality}  \\ &= k! \cdot \prod_{\mathrm{bins} \ i} \frac{1}{{k_i}!} \left(\frac{\lambda_i}{\lambda}\right)^{k}  \label{eq:mn_ratio_form_standard}
\end{align}
It turns out that the same relation does not hold anymore in the finite-sample limit. If we just replace the Poisson and Multinomial factors in eq. (\ref{eq:poisson_mn_equality}) with the finite-sample expressions from section \ref{sec:poisson} and section \ref{sec:multinomial}, we can check numerically that equality is slightly broken - with the exception of all weights in all bins being equal, i.e. when the Multinomial expression is the Dirichlet-multinomial distribution (see eq. \ref{eq:multinomial_dirichlet1}). However, we can still try to perform such a construction and check what the outcome actually corresponds to. Interestingly, the results are slightly different probability distributions than the "standard" finite-sample counterparts, and some of their properties are described in the following. For all expressions in this section, we use the unique Prior, i.e. $\alpha=0$, as it seems to be the only sensible choice when expressions with different bin definitions are combined. 

\subsection{"Product" construction for a likelihood of Poisson type}

\begin{figure}%
	\centering
	\subfloat[Standard Poisson]{\scalebox{0.9}{\includegraphics{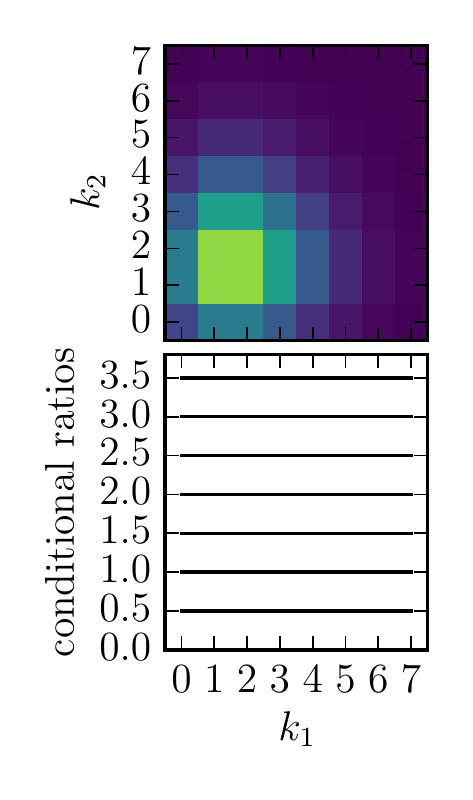}}}
	\subfloat[Finite-sample Poisson]{\scalebox{0.9}{\includegraphics{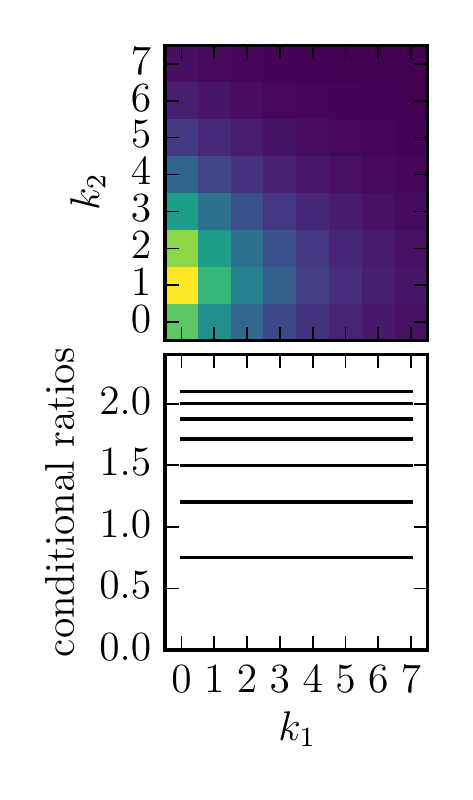}}}
	\subfloat[Product construction]{\scalebox{0.9}{\includegraphics{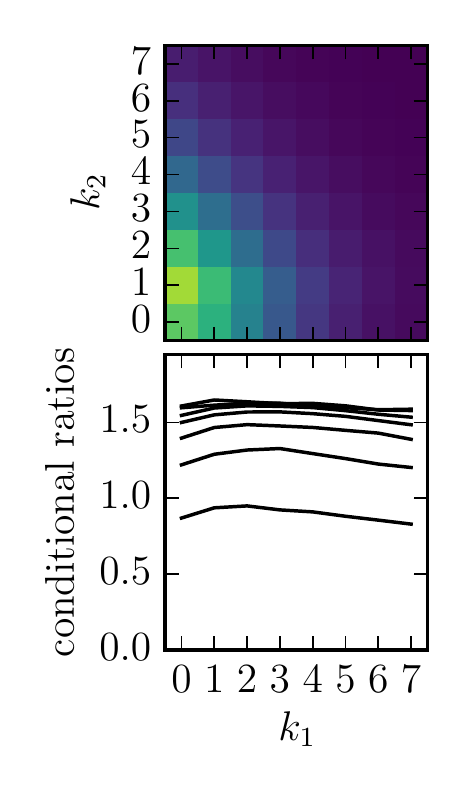}}}
	\caption{Different Poisson constructions for two bins compared by their PDFs of the counts $k_1$ and $k_2$. The first bin contains one weight with magnitude $2$, the second bin contains $4$ weights with magnitude $0.5$ each, so the sum of weights in both bins is equal. The upper plot shows the 2-d PDF (z-axis omitted, equal scale), the lower plot shows the ratio of conditional distributions $P(k_1; k_2 = i)/P(k_1; k_2 = i + 1)$ for various $i$. The three Poisson-type formulas used for (a),(b),(c) are eq. (\ref{eq:poisson_with_sumweights}), eq. (\ref{eq:equal_form_one}) and eq. (\ref{eq:product_construction_equal}), respectively. The non-standard constructions use the unique Prior.} \label{fig:multivariate_construction}
\end{figure}

In this section, we take eq. (\ref{eq:poisson_mn_equality}) as inspiration and multiply a global finite-sample Poisson factor with a finite-sample Multinomial expression and then observe how the end result behaves. For equal weights per bin, we multiply eq. (\ref{eq:finite_multinomial_equal}) with eq. (\ref{eq:equal_form_one}) to obtain

\begin{align}
L_{\mathrm{\mathbf{P}, product, equal}} &= L_{\mathrm{\mathbf{MN}, equal}} \cdot L_{\mathrm{\mathbf{P}, total, equal}} \label{eq:product_construction_equal}
\end{align}
and for general weights, we multiply eq. (\ref{eq:finite_mnomial_general}) with eq. (\ref{eq:finite_poisson_finitesum}) to obtain

\begin{align}
L_{\mathrm{\mathbf{P}, product, gen.}} &= L_{\mathrm{\mathbf{MN}, gen.}} \cdot L_{\mathrm{\mathbf{P}, total, gen.}} \label{eq:product_construction_general}
\end{align}
where for simplicity we do not write out the total expressions, as no insightful simplification is possible at this point. It turns out these "product" expressions are multivariate probability distributions in $\bm{k}$ with a certain correlation structure imposed from the difference of the weights. A simple example with two bins is shown in figure \ref{fig:multivariate_construction}.
It depicts the difference between the standard Poisson likelihood (eq. \ref{eq:poisson_with_sumweights}), the finite-sample extension (eq. \ref{eq:equal_form_one}), and the product construction (eq. \ref{eq:product_construction_equal}) for an artificial example with two bins: the first bin contains one weight with magnitude $2$, the second bin four weights with magnitude $0.5$ each, so the sum of weights is the same in each bin. The standard Poisson PDF in figure \ref{fig:multivariate_construction}-(a) is symmetric, and does not know about the different weight distributions in the two bins. Also, the conditionals are flat, since the two Poisson factors for each bin are just multiplied with each other. The PDF for the finite-sample extension in figure \ref{fig:multivariate_construction}-(b) has more confidence in the second bin, which contains more weights, and the distribution becomes asymmetric. The conditional distributions are still flat, since by construction the PDF is still a product of individual finite-sample Poisson factors which results in independently distributed random variables. The PDF for the product construction in figure \ref{fig:multivariate_construction}-(c) is again asymmetric, but not independent anymore, as can be seen from the skewed ratios of conditional distributions. This dependence, or correlation, comes from the difference of the weights. If the weights were changed to be equal in all bins, the resulting distributions for (b) and (c) would be similar. Whether such a correlation is desired in practice remains to be seen, and would probably depend on the problem. The result implies that there are at least two principle ways to write a Poisson likelihood in the finite-sample limit - once as derived in section \ref{sec:poisson} by multiplying individual finite-sample Poisson factors for each bin, and once via a product construction that may contain multivariate correlation structure when the weights differ. When all weights in all bins are identical, which includes the limit of infinite statistics, both agree with each other.

\subsection{"Ratio" construction for a likelihood of multinomial type}
\label{sec:ratio_constructions}

For "ratio" constructions, i.e. forming a multinomial-like expression from Poisson factors, we can derive some more rigorous and practical results. We start by writing the analogue of eq. (\ref{eq:mn_ratio_form_standard}) for finite Monte Carlo events assuming equal weights per bin. Using eq. (\ref{eq:equal_form_one}) and eq. (\ref{eq:laplace_type_lauricella_replaced}), the resulting expression looks like

\begin{align}
L_{\mathrm{\textbf{MN}, ratio, eq.}}&=\frac{\prod_i L_{\mathrm{\mathbf{P},equal,i}}}{L_{\mathbf{P},general}} \\
&\begin{aligned}
&= \prod_{\mathrm{bins} \ i} \frac{(1/w_i)^{k_{mc,i}}  \cdot (k_i+k_{mc,i} - 1)!}{  (k_{mc,i} - 1)! k_i!\cdot(1+1/w_i)^{k_i+k_{mc,i}}} \Bigg/ \\ & \Bigg[ \frac{(k_{mc}+k)!}{(k_{mc}-1)! k!} \cdot \left(\frac{1}{w_N}\right)^{k_{mc}} \cdot \left(\frac{1}{1+1/w_N}\right)^{k_{mc}+k} \cdot \left( \prod_{j=1}^{N} \left(\frac{w_N}{w_j}\right)^{k_{mc,j}} \right) \\ & \cdot F_D(k_{mc}+k;\bm{b^*};k_{mc};\bm{z^{**}})  \Bigg]
\end{aligned} \\
&= \mathrm{\textbf{DM}}(\bm{k}; \bm{k_{mc}}) \cdot \left(\prod_i \left(\frac{1+1/w_N}{1+1/w_i} \right)^{k_{mc,i}+k_i} \right) \cdot \frac{1}{F_D(k_{mc}+k;\bm{b^*};k_{mc};\bm{z^{**}})} \label{eq:mn_ratio_form_finite}
\end{align}
, consisting of a Dirichlet-multinomial factor, a factor depending on the weights, and the inverse of $F_D$ with specific arguments $b*_i=k_{mc,i}$ and $z^{**}_i = 1-\frac{1+1/w_i}{1+1/w_N} \ (i=1 \ldots N-1)$ where $w_N$ is the smallest weight in all bins. Since the Dirichlet-multinomial distribution $\mathrm{\textbf{DM}}$ is a proper probability density in $\textbf{k}$, i.e. the vector of observed counts in the individual bins, and eq. (\ref{eq:mn_ratio_form_finite}) is proportional to $\mathrm{\textbf{DM}}$, we can write

\begin{align}
L_{\mathrm{\textbf{MN}, ratio, eq.}} &=  \mathrm{\textbf{DM}}(\bm{k};\bm{k_{mc}}) \cdot \frac{\prod_i \left(\frac{1+1/w_N}{1+1/w_i} \right)^{k_{mc,i}+k_i} }{F_D(k_{mc}+k;\bm{b^*};k_{mc};\bm{z^{**}})} \label{eq:ratio_poisson_probdist_poisson} \\
&= \mathrm{\textbf{DM}}(\bm{k};\bm{k_{mc}}) \cdot \frac{C(\bm{k})   }{\stackrel[\sum_i k_{*,i}=k, \ k_{*,i} \geq 0 ]{}{\sum  } \mathrm{\textbf{DM}}(\bm{k_*};\bm{k_{mc}}) \cdot C(\bm{k_*})} \label{eq:ratio_poisson_probdist}
\end{align}
From this construction, we see that $L_{\textbf{MN},\mathrm{ratio, eq.}}$ is a probability distribution if $C(\bm{k})=\prod_i \left(\frac{1+1/w_N}{1+1/w_i} \right)^{k_{mc,i}+k_i}$ and $F_D(k_{mc}+k;\bm{b^*};k_{mc};\bm{z^{**}})=\stackrel[\sum_i k_{i}=k, \ k_{i} \geq 0]{}{\sum  } \mathrm{\textbf{DM}}(\bm{k};\bm{k_{mc}}) \cdot C(\bm{k})$, since the "artificially" constructed denominator acts as a normalizing constant. Exactly this has been shown in eq. (\ref{eq:multinomial_rep_for_fd}), which confirms that the above ratio is indeed a probability distribution. More generally, we could have used any definition of $z_j^{**}=1-\frac{c_1+c_2/w_j}{c_1+c_2/w_N}$, and the same arguments still lead to a probability distribution. The choice of $c_1=0$ and $c_2=1$, for example, corresponds to the finite-sample limit of $K \cdot \frac{\prod_i E[\lambda_i/k_i!]}{E[\lambda/k!]}$, the ratio of expectation values without the respective exponential factors in the numerator and denominator. 
For a general weight distribution, the corresponding ratio representation similar to eq. (\ref{eq:mn_ratio_form_standard}) and eq. (\ref{eq:ratio_poisson_probdist_poisson}) results to be

\begin{align}
L_{\mathrm{\textbf{MN}, ratio, gen.}} &= \mathrm{\textbf{DM}}(\bm{k};\bm{k_{mc}}) \cdot \frac{\prod_i  \left(\frac{1+1/w_N}{1+1/w_{N,i}} \right)^{k_{mc,i}+k_i} F_D(k_{mc,i}+k_i;\bm{b_i^*};k_{mc,i};\bm{z_i^{**}}) }{F_D(k_{mc}+k;\bm{b^*};k_{mc};\bm{z^{**}})} \label{eq:ratio_general}
\end{align}
, where $w_N$ is the smallest weight in all bins and $w_{N,i}$ is the smallest weight in bin $i$. This expression cannot be shown to be a probability distribution in the same way as before, but a numerical check shows that it is. Using eq. (\ref{eq:ratio_poisson_probdist}), i.e. writing the expression in terms of a Dirichlet-multinomial factor and an unknown $C(\bm{k})$, this implies that one can write $F_D$ as a nested sum over other $F_D$'s with less parameters by comparison of the denominators in eq. (\ref{eq:ratio_poisson_probdist}) and eq. (\ref{eq:ratio_general}). Since the form of $C(\bm{k})$ depends on the partition of weighted Monte Carlo events in the chosen binning, there is one distinct representation per possible weight partition for $F_D$. It would be good to have a mathematical foundation of these representations.

We can now compare these "ratio constructions"  with the standard multinomial finite-sample extension (eq. \ref{eq:finite_multinomial_equal}).  For the ratio construction, $F_D(a;\bm{b};c;\bm{z})$ appears with $a>c$, while in the standard multinomial extension $F_D(a;\bm{b};c;\bm{z})$ appears in the numerator with $a<c$, at least for equal weights per bin. The latter $F_D$ has an exact representation involving logarithms \cite{Tan2005}, which shows it can likely not be brought to similar analytic form. This means there are at least two ways of writing a finite-sample extension of the multinomial likelihood: once as a ratio of finite-sample Poisson expressions, and once directly derived from the expectation value of the multinomial likelihood as shown in section \ref{sec:multinomial}.  Only when all weights in all bins are the same, they are equivalent. In the limit of infinite statistics, they both converge to the multinomial likelihood. 

\begin{figure}%
	\centering
	{\includegraphics{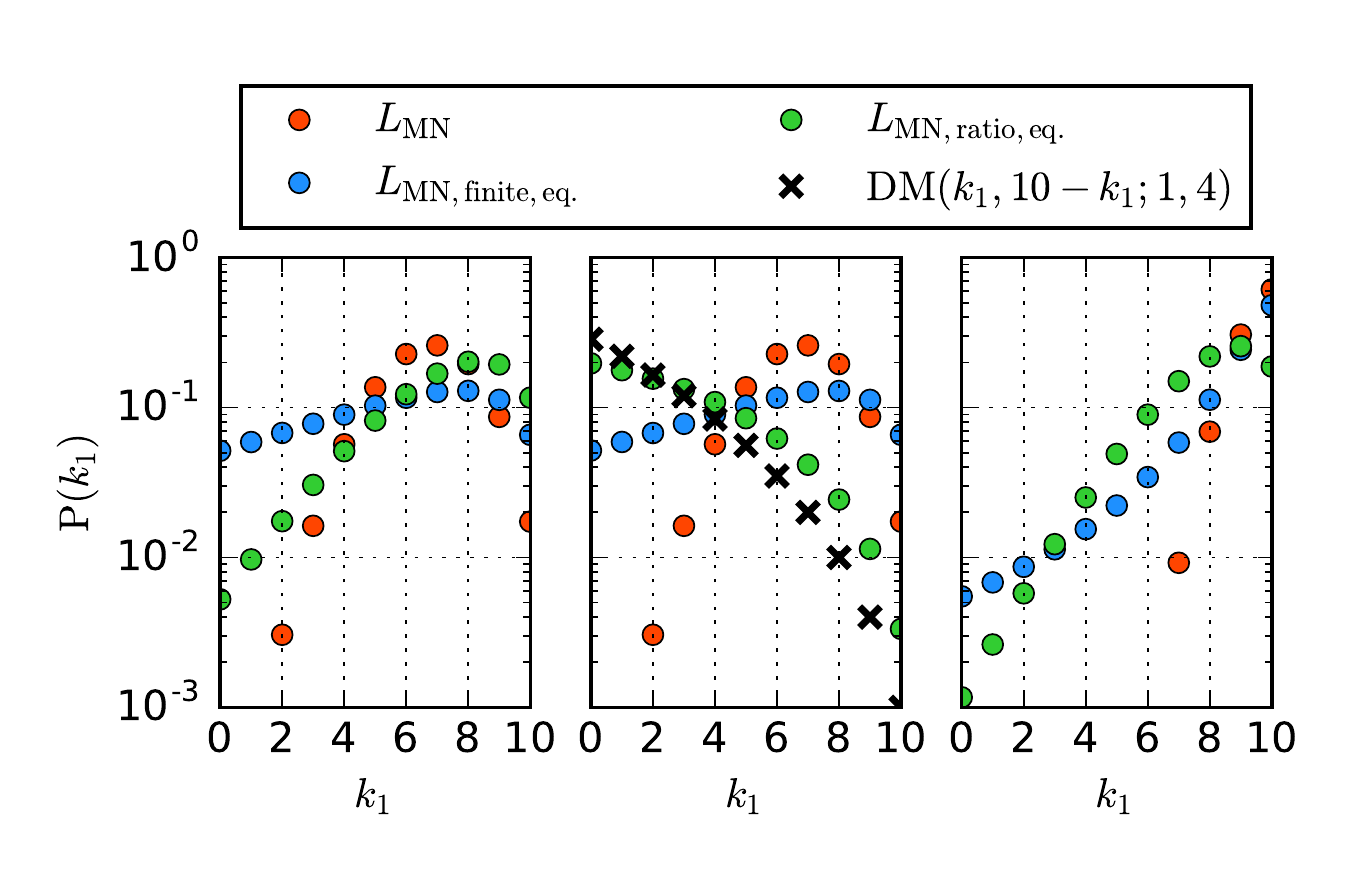}}
	\caption{Comparison of multinomial-like formulas in three artifical situations with two bins and the following MC weight structure: $[1\cdot 4, 4 \cdot 0.5]$ (left), $[1\cdot 40, 4 \cdot 5]$ (center), $[1\cdot 40, 4 \cdot 0.5]$ (right). The counts in the first bin are denoted as $k_1$, and the total number of counts are $N=10$.} \label{fig:ratio_multinomial}
\end{figure}

There is another important property of ratio constructions for $c_0\neq0$, which applies in eq. \ref{eq:ratio_poisson_probdist_poisson}. The behavior depends on the overall scale of the weights, since the absolute weight scale does not cancel out. This  happens especially pronounced when the individual weights are larger than unity, since then $c_0$ dominates the expression. The behavior is illustrated in figure \ref{fig:ratio_multinomial},
which compares different PDFs of multinomial type for two bins, i.e. in the binomial setting, given a total number of events $N=10$. In the first plot, the first bin contains one weight with magnitude 4, the second bin 4 entries with magnitude $0.5$ each, resulting in a sum of $2$. The total weight structure can shortly be summarized as $[1\cdot 4, 4 \cdot 0.5]$. One therefore traditionally expects a peak at roughly $k_1 \approx 0.66 \cdot N$, since the ratio of the sum of weights is $2/1$. This is observed for the standard multinomial likelihood (red). The finite-sample extension (blue) includes the uncertainty of the finite event Monte Carlo and has a wider shape, especially for low counts. The ratio expression (green) gives less importance to MC events with weight larger than unity, and therefore is biased towards high values of $k_1$. In the second column, all weights are multiplied by $10$, giving a weight structure of $[1\cdot 40, 4 \cdot 5]$. The blue and red curves are unchanged, since in the standard multinomial and finite-sample multinomial PDF the overall weights do not play a role. The ratio-construction (green) changes completely, approaching a Dirichlet-multinomial distribution. For even larger overall weights the distribution eventually matches the DM distribution exactly. This means, for weights larger than unity, the events become asymptotically equally important and their weights meaningless. In the last column the weight distribution is changed to $[1\cdot 40, 4 \cdot 0.5]$, i.e. this time only the first bin weight gets upscaled by $10$. This changes the traditional multinomial (red) and finite-sample multinomial (blue) construction towards higher values for $k_1$, as expected since the first weight has a much larger overall share. For the standard multinomial PDF this is dramatic, and low $k_1$ are strongly excluded. In the ratio construction (green), on the other hand, the PDF is not too different from the ratio construction in the first plot, since the importance of weights larger than unity is reduced. In practice, only a few MC events are usually larger than unity, and reducing their importance might be desired behavior. This has to be studied with care, and probably depends on the application.

\section{Toy example: determining the normalization of a peak in an energy spectrum}

\begin{figure}[ht]%
	\centering
	\subfloat[Likelihood scan -  268 MC events]{\scalebox{0.9}{\includegraphics{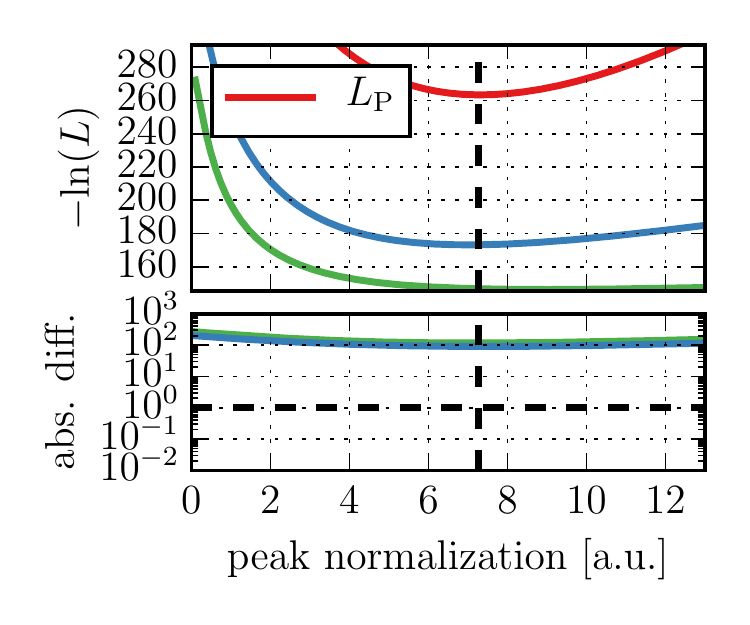}}} \subfloat[Observable binning - 268 MC events]{\raisebox{0.25\height}{\includegraphics[width=0.5\textwidth]{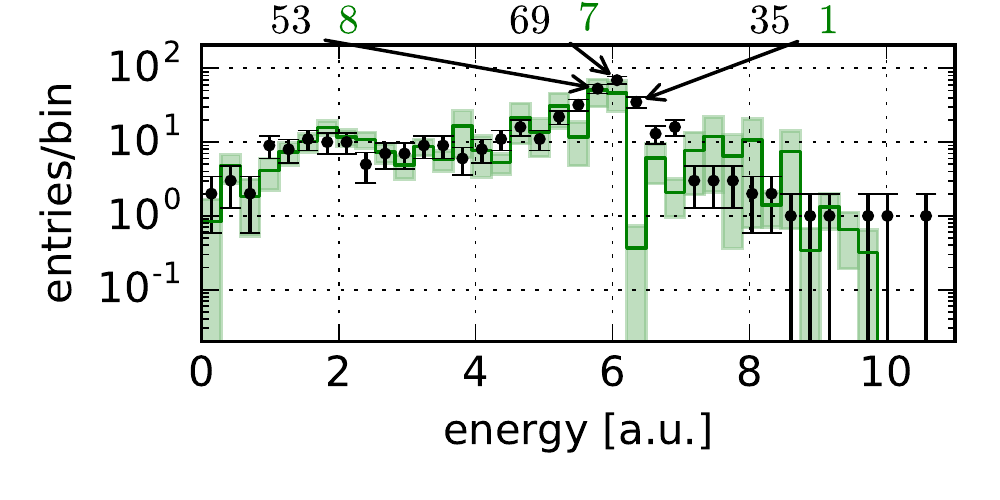}}} \\
	\subfloat[Likelihood scan - 2682 MC events]{\scalebox{0.9}{\includegraphics{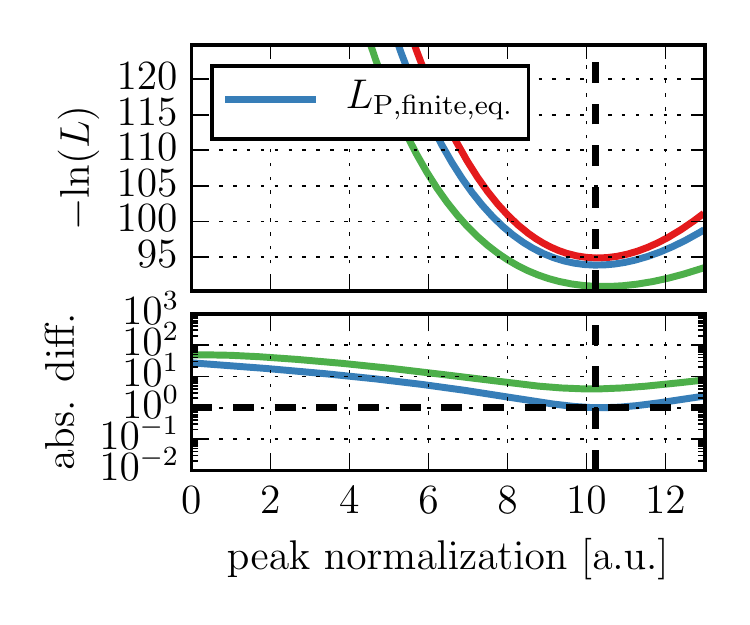}}} \subfloat[Observable binning - 2682 MC events]{\raisebox{0.25\height}{\includegraphics[width=0.5\textwidth]{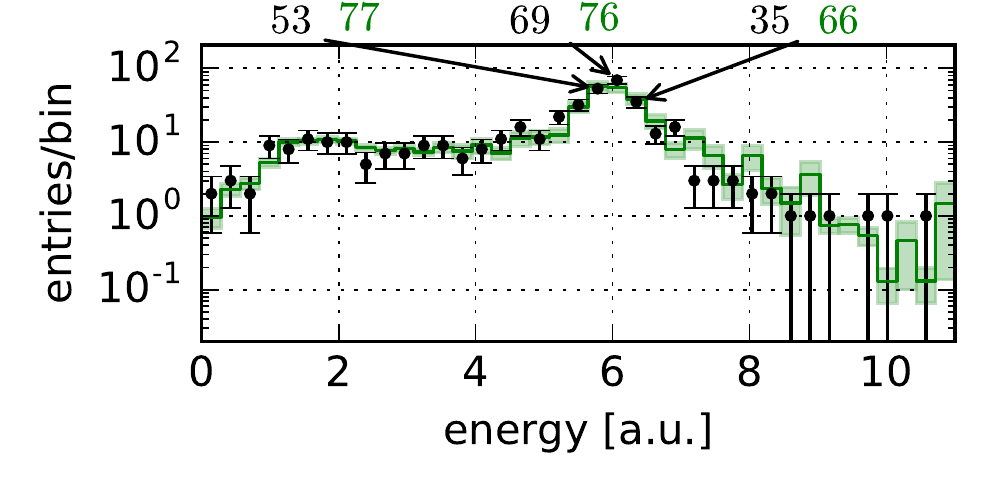}}} \\
	\subfloat[Likelihood scan -  26821 MC events]{\scalebox{0.9}{\includegraphics{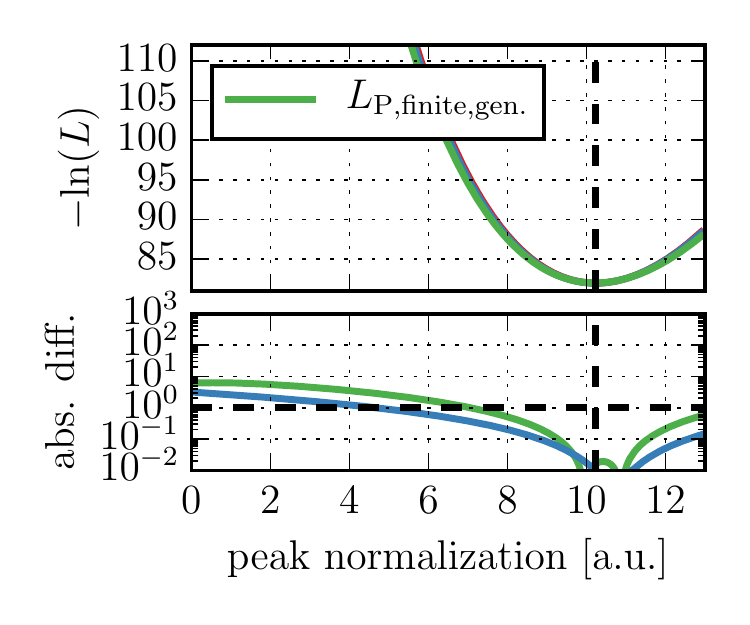}}} \subfloat[Observable binning - 26821 MC events]{\raisebox{0.25\height}{\includegraphics[width=0.5\textwidth]{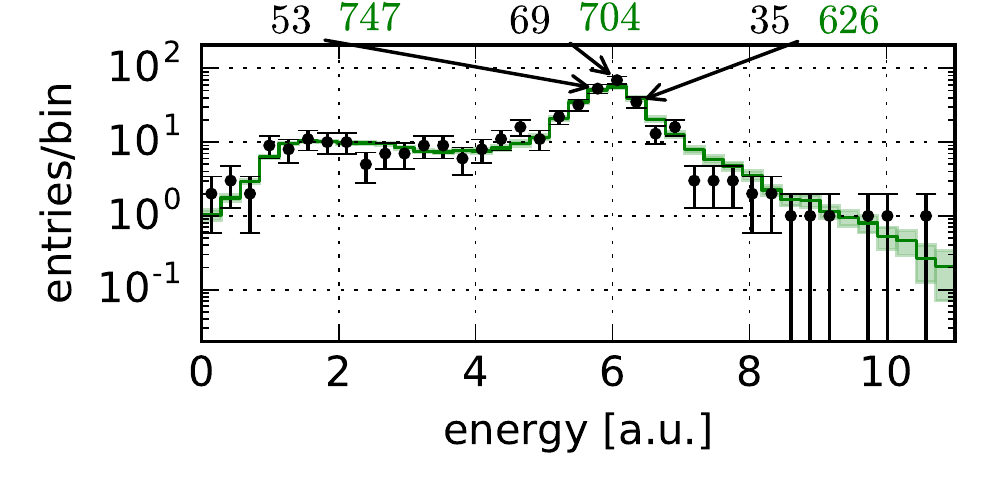}}} \\
	\caption{Likelihood scans (left column) and respective observable space (right column) for different amounts of MC events. The data is the same in all figures.  Left column: The standard Poisson calculation is shown in red (eq. \ref{eq:poisson_with_sumweights}), the equal-weight finite-sample expression in blue (eq. \ref{eq:equal_form_two}), and the general expression in green (eq. \ref{eq:finite_poisson_finitesum}). The vertical dotted line indicates the minimum of the standard Poisson likelihood. The lower part shows the absolute difference of $-\mathrm{ln}(L)$ to $L_{\mathrm{P}}$ (red). Right column: Data is shown in black and MC (sum of weights) in green. The arrows point to the most relevant bins for the peak determination, and show their respective data and MC counts.   } \label{fig:llh_scan_and_observable_space}
\end{figure}

The behavior of the modified Poisson likelihood is demonstrated with a toy study: a simulation of a falling energy spectrum with an additional peak with a certain normalization. The aim is to measure the normalization of the peak. The position and width of the peak are fixed, and a likelihood scan is performed in one dimension, taking into account effects of the artificial detector response. In this example, these include energy smearing and energy-dependent detection efficiency. Figure \ref{fig:llh_scan_and_observable_space} demonstrates the behavior of increasing MC statistics for the scan and for the observable space. For low statistics, the absolute log-likelihood values between the three approaches differ drastically, as do the position and width of likelihood-based confidence intervals. The width of the confidence intervals including the MC uncertainty is always larger. With increased MC statistics, the curves approach each other, and are eventually indistinguishable. The corresponding observed "energy" distribution gives an idea of the MC fluctuations in the individual bins.

\begin{figure}[ht]%
	\centering
	\subfloat[Prior comparison - 40 bins total]{\scalebox{0.75}{\includegraphics{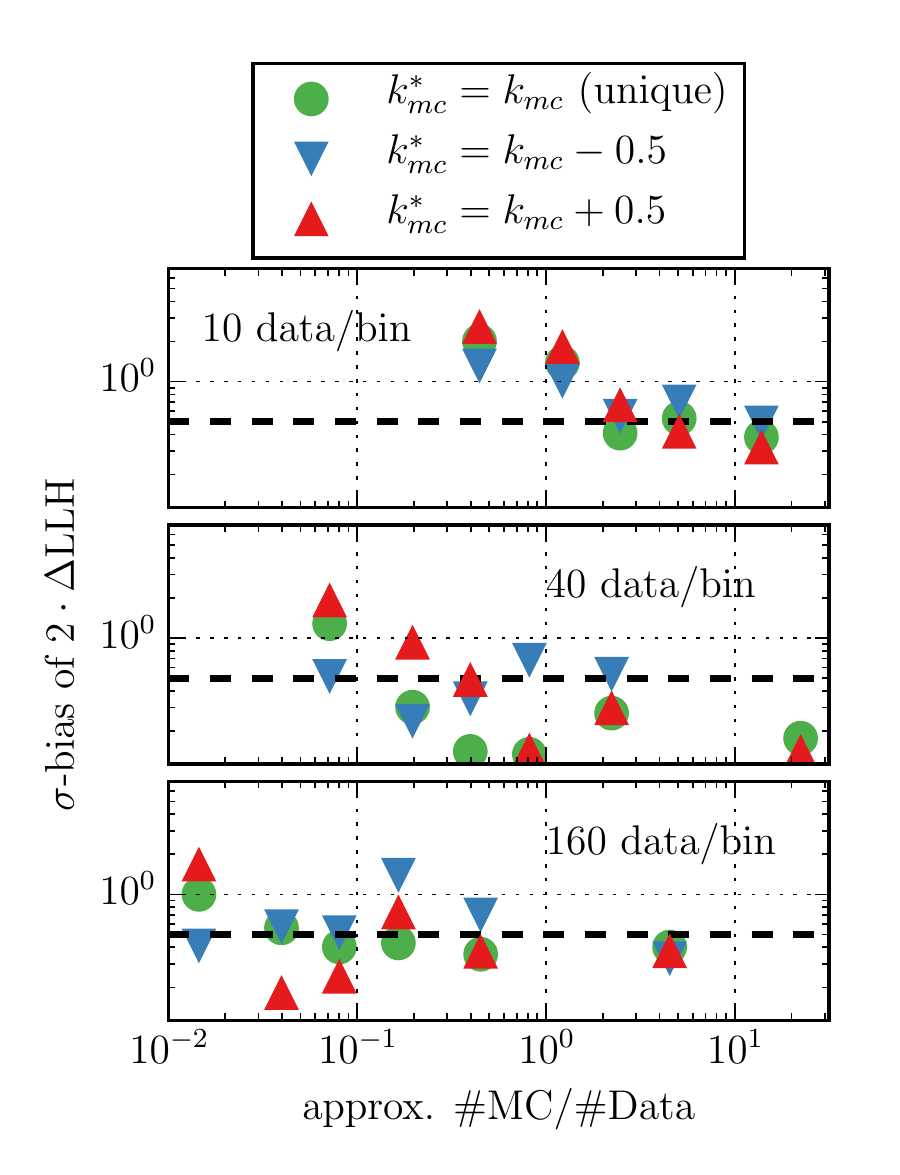}}} 
	\subfloat[Prior comparison - 200 bins total]{\scalebox{0.75}{\includegraphics{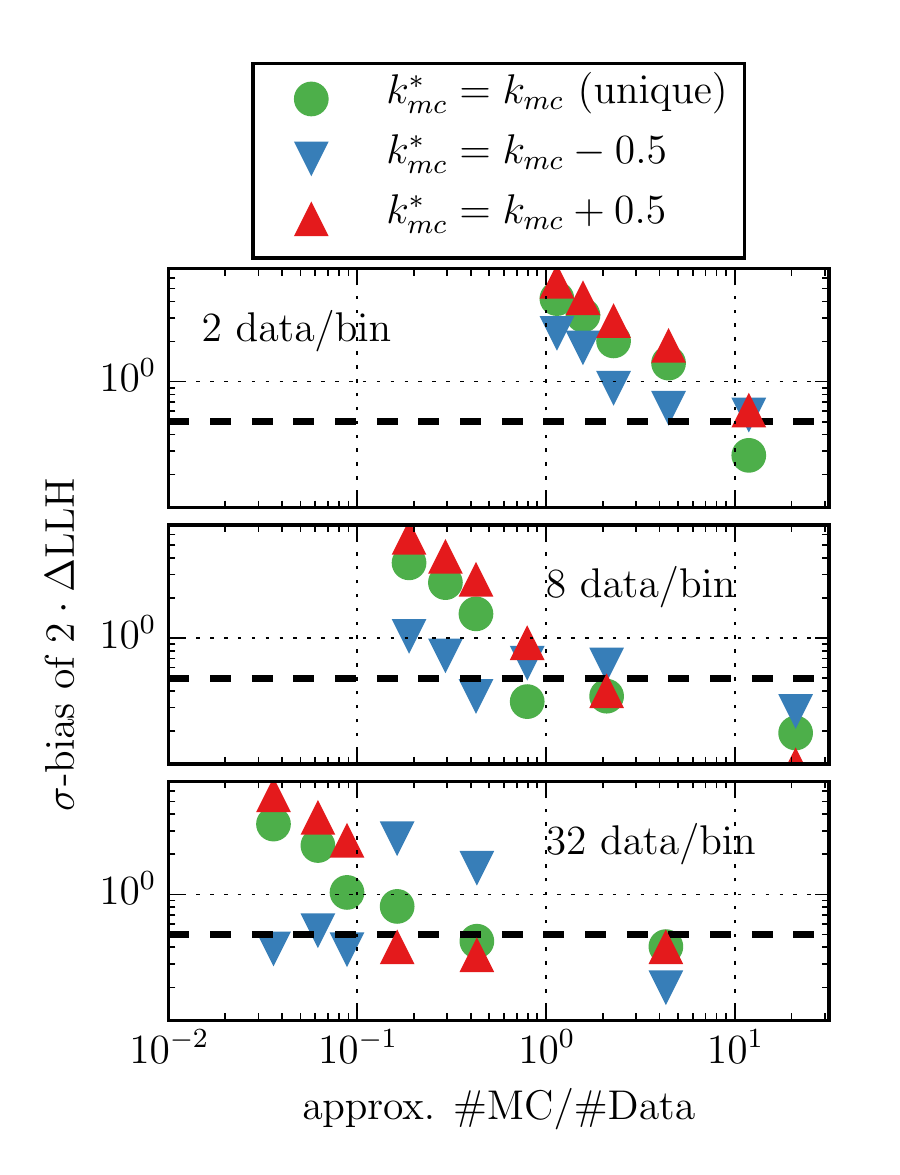}}} \\
	\subfloat[Method comparison - 40 bins total]{\scalebox{0.75}{\includegraphics{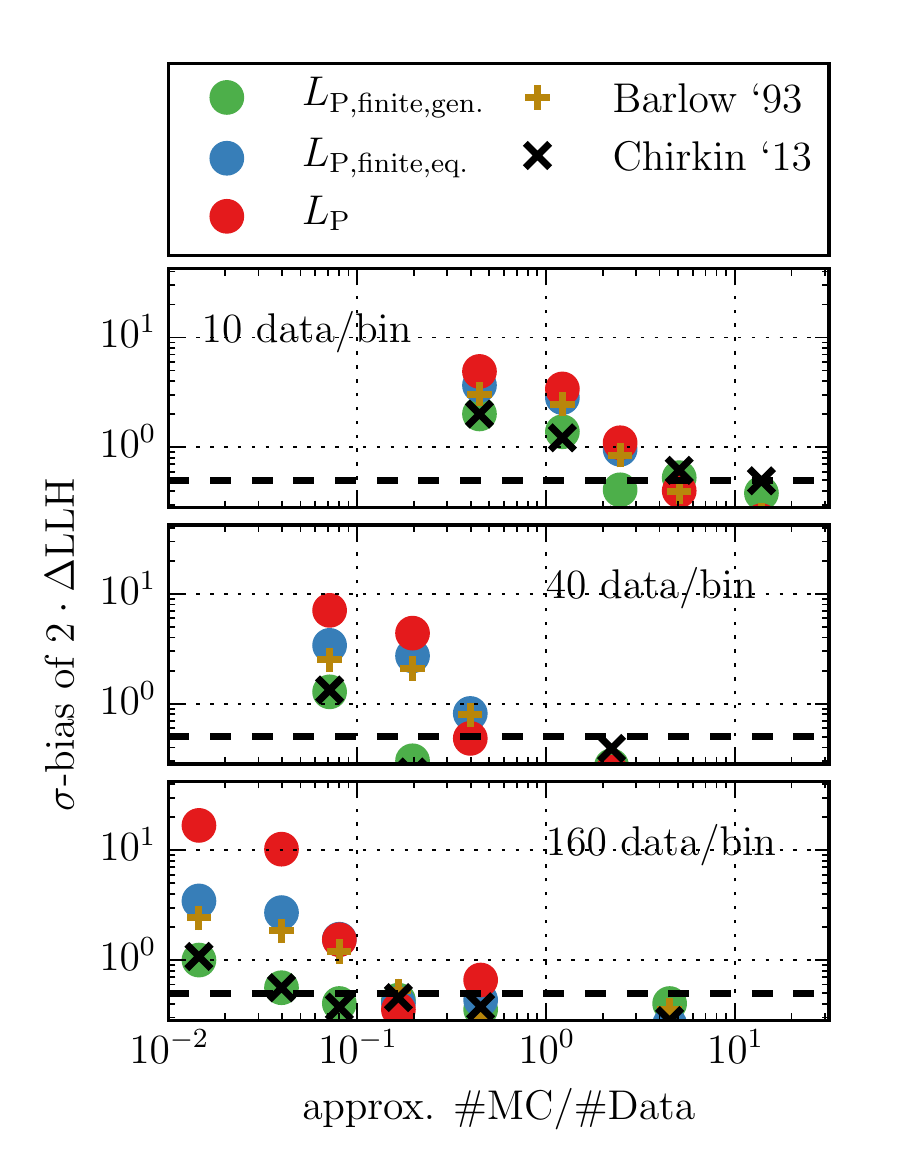}}} \subfloat[Method comparison - 200 bins  total]{\scalebox{0.75}{\includegraphics{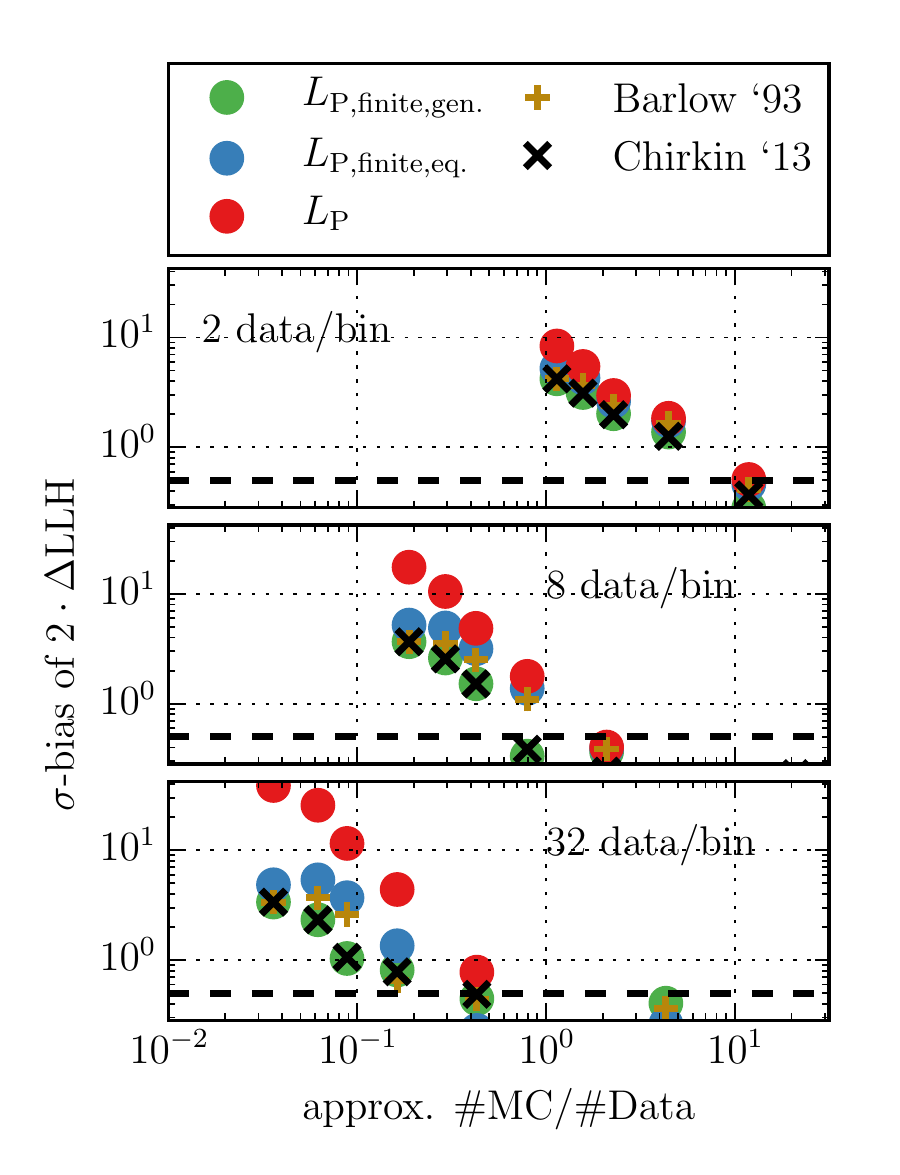}}} \\
	\caption{Bias (difference) of $2\cdot \Delta LLH$ between a given formula and the standard Poisson likelihood. The LLH-ratio is calculated with respect to the true parameter value. The x-axis shows the approximate ratio of MC to data in the relevant fit region (around peak, see fig. \ref{fig:llh_scan_and_observable_space}). The absolute no. of data events per bin in this region is given in each plot. The thick horizontal dashed line indicates that values below $\approx 0.5  \sigma$ should not be used to draw conclusions due to wrong asymptotic assumptions for the standard Poisson likelihood (see text).} \label{fig:llh_bias}
\end{figure}

The ability to capture the uncertainty from finite Monte Carlo events can be quantified with a certain bias definition, the difference of twice the log-likelihood-ratio ($2\cdot \Delta LLH$) using a given likelihood formula and the standard Poisson likelihood for infinite MC statistics. It is convenient to express this  bias in "$\sigma$-equivalents", i.e. how many $\sigma$ is the result systematically off by not having enough statistics.
Figure \ref{fig:llh_bias} shows this bias for different simulated live times and different binning schemes, but using exactly the same data. Bias values below 0.5 should be ignored, since fluctuations of this magnitude just randomly occur because the Poisson formula is used as the "infinite MC data" reference, and infinite statistics are not reached in practice. The 40-bins scheme is the same is in figure \ref{fig:llh_scan_and_observable_space}.
Figures \ref{fig:llh_bias} (a) and (b) compare the bias for different Prior choices. The compared values are $\alpha=0$ (unique Prior) and two slightly varied values $\alpha=-0.5$ and $\alpha=0.5$, where $\alpha$ has been defined in eq. (\ref{eq:convolution_prior_parameter}). The value $\alpha=0.5$ is motivated by a common choice which is scale invariant under re-parametrizations (Jeffreys Prior) \cite{Jeffreys1946} and is proportional to $~\lambda^{-0.5}$ for Poisson rate inference. Using $\alpha=0.5$ corresponds to the Jeffreys Prior when all weights are equal and the solution is afterwards scaled by a scaling factor proprtional to the weight (see for example \cite{Aggarwal2012}). The value $\alpha=-0.5$ is chosen as the opposite for simplicity. The unique Prior has generally less bias than $\alpha=0.5$, but seems to be worse than $\alpha=-0.5$ for low statistics. However, the bias for $\alpha=-0.5$ has non-monotic behavior and rises for intermediate Monte Carlo statistics substantially, independently of the binning. It is not clear why that happens, but certainly it is undesired behavior. The unique Prior therefore seems to be a preferred choice. We use it in figure \ref{fig:llh_bias} (c)+(d) as well.
 
Figure \ref{fig:llh_bias} (c)+(d) compare the standard Poisson likelihood, the equal-weights extension (eq. \ref{eq:equal_form_one}) and general-weights extension (eq. \ref{eq:finite_poisson_finitesum}) with two previous methods found in the literature. These are the methods by Barlow et al. \cite{Barlow1993} and Chirkin \cite{Chirkin2013}, which essentially optimize for nuisance parameters instead of integrating the distribution of $\lambda$. In the method by Barlow et al. \cite{Barlow1993}, one has to specify a certain number of "MC sources". We choose one source, since it is ultimately not always clear how the Monte Carlo is generated. The peak in the toy-MC could for example come from the same dataset, were events are just re-weighted, or from a second simulation dataset which is added to the baseline. In the actual calculation, all weights from a given MC dataset are averaged. This is in contrast to \cite{Chirkin2013}, which uses quite similar formulas, but limits the number of MC sources to a single dataset, while the whole weight structure is taken into account. It would have been interesting to also include the method proposed in \cite{Bohm2014} based on an approximation of the compound Poisson distribution of weight distributions, i.e. a sum of weight random variables where the number of convolutions is Poisson distributed. Our Poisson Ansatz can be viewed as an analytic approximation of that distribution, with subsequent integration over $\lambda$, so we would expect the results to be quite comparable.

Looking at figure \ref{fig:llh_bias} (c)+(d), the lowest bias comes from the general-weights likelihood (eq. \ref{eq:finite_poisson_finitesum}) and the method proposed in \cite{Chirkin2013}, whose results are practically identical. Next comes the method proposed in \cite{Barlow1993}, which is not surprising since weights are averaged in the process. It might be that this method would give better results, if the bump was modeled via a second MC dataset - this however would require an additional nuisance parameter, and again it is not always clear that such a situation exists. Slightly worse still is the performance of the equal-weights formula (eq. \ref{eq:equal_form_one}) using the average weight per bin. This might indicate the result is slightly more biased by performing a "wrong" integral than by performing a "wrong" optimization - "wrong" here means using the average weight in a bin.

Surprisingly, we can reach a greater bias reduction as the number of data increases. For 2 data per bin, for example, the usual rule of thumb \cite{Barlow1993} of 10 times as much MC as data is necessary in order to be unbiased, even using a modified likelihood. In the other extreme, for 250 data events / bin, the result is essentially unbiased down to a tenth of the simulated live time. The relative bias reduction potential of the modified likelihoods is therefore larger, the more overall live time is analyzed, i.e. the more data events are present. We are not aware of any previous systematic study of this behavior. The remaining bias that is visible for very low counts possible comes in part from the uncertainty due to the MC sampling realization, which is not taken into account (see section \ref{sec:the_problem_definition}).

\section{Conclusion}

We have shown that there is an analytic way to generalize the Poisson and Multinomial PDF to handle Monte Carlo-based finite-sample uncertainty via marginalization with a suitable probability distribution $P$. For the Poisson likelihood, $P=P(\lambda)$ is a convolution of general gamma distributions. For the multinomial likelihood, in the special case of having equal weights per bin, $P=P(p_1, \ldots, p_{N-1})$ is a scaled Dirichlet distribution. The parameters of these functions are fixed from consistency conditions up to a parameter $\alpha$. The case $\alpha=0$ is a special choice, which is motivated by the additional consistency condition that the Prior does not depend on the number of bins or number of Monte Carlo events in a bin, which we call "unique". The resulting expressions are PDFs in the data, and therefore can be used as likelihood replacements in any likelihood analysis, typically in forward-folding likelihood fits.

This is the first approach that not only applies to the Poisson, but also to the multinomial likelihood, and it reduces to the respective standard expression in the limit of infinite statistics. Since the multinomial likelihood is proportional to an unbinned likelihood that uses sample-derived PDFs, the method can also be used to incorporate the finite-sample uncertainty in unbinned likelihood fits where PDFs are approximated by Monte Carlo data. This is for example the case in high-energy neutrino point source searches, where parts of the total PDF are modeled with MC-derived binned PDFs \cite{Aartsen2014}.

The calculations involve hypergeometric functions,
most notably the fourth Lauricella function $F_D$. For $F_D(a;\bm{b};c;\bm{z})$ with $a >c$ and $a-c$ being integer, we find a new exact finite-sum representation. This representation allows to calculate the finite-sample Poisson likelihood for general weights orders of magnitude faster than numerical integration, which makes it usable in practice. Since $F_D$ can be related to the Dirichlet average $R_n$, the result can also be used to write down compact finite-sum expressions for the probability generating function of the Dirichlet-multinomial probability distribution, the divided difference of monomials, or the calculation of the moments of univariate B-splines with arbitrary knot positions (see \ref{appendix:pgf_and_others}).

In general, all new formulas come 2-fold: a formula for general weights, and a second simpler one derived for equal weights per bin, which can also be used in the general case as an approximate formula by plugging in the average weight per bin. In addition, we describe non-standard constructions that are motivated by well-known relationships of Poisson and multinomial factors in the asymptotic limit of infinite statistics, which in the finite-sample case leads to different expressions. The first is a "product" construction that mimics Poisson behavior, but has multivariate correlation between bins. The second is a "ratio" construction that mimics multinomial behavior, but has an overall normalization dependence. The usage of these different constructions might be desired in certain situations, but has to be studied in detail. The "ratio" construction certainly offers computational advantages compared to the standard multinomial finite-sample expression.

In the final section we demonstrate the bias-reduction for parameter estimation using the modified Poisson likelihood formula with a typical toy-MC problem, where the normalization of a peak on a falling energy spectrum is determined with a likelihood scan. We quantify the results with a $\sigma$-equivalent bias with respect to the LLH-ratio one would have obtained if infinite statistics were available. First, we show that the unique Prior ($\alpha=0$) performs better than slight $\alpha$ variations away from zero, in particular also better than Jeffreys Prior, in terms of monotonic behavior towards larger MC statistics. Then we compare the new formulas using the unique Prior with other approaches in the literature. The general formula gives practically indistinguishable results to the approach derived in \textit{Chirkin} \cite{Chirkin2013}. The method by \textit{Barlow et al.}\cite{Barlow1993} has a little larger bias, which is not too surprising since it depends on the average weight per bin. Using the average weight per bin for the equal-weights formula is a little worse still. Given that in these other approaches a nuisance parameter is optimized, instead of marginalized, one can think of the methodology here as a probabilistic counterpart to these Frequentist methods. The method can also be interpreted as an analytic approximation to the compound Poisson distribution advocated for in \cite{Bohm2014}, which is then integrated over with an additional Poisson factor. 

The toy-MC further reveals that parameter estimation for low-count analyses requires the well-known rule of thumb that roughly $10$ times the amount of Monte Carlo data is desirable to get rid of Monte-Carlo related effects. Interestingly, the more data is used in the measurement, the more this requirement is relaxed, at least if bias is concerned. In some of the shown examples only a tenth of the actual live time seems to be sufficient for the Monte Carlo simulation to obtain an unbiased result with respect to the infinite statistics limit. The remaining bias for very low data counts might come from the residual uncertainty due to the actual MC sampling step that is still neglected (see section \ref{sec:the_problem_definition}). It should be mentioned that all bias-related studies could differ depending on the type of parameter that is being studied.

In a sense, one should think of the probability distributions described in this paper as more precise probability distributions for bin counts in the presence of Monte Carlo-based expectations. If traditional probability distributions (i.e. Poisson and multinomial) are used, these new distributions provide the means to check that the Monte Carlo sample size has no effect - for example via a comparison of their absolute likelihood values or a crosscheck likelihood scan, similar to the discussed toy example. For the case of Poisson evaluation with general weights, and when the absolute likelihood value is unimportant, which it often is, the method introduced in \cite{Chirkin2013} is a good alternative that seems to have equal precision, but is usually faster. However, we recommend to try the new formulas for the case of multinomial evaluation, for unbinned fits with MC-derived PDFs, for MC estimates with equal weights per bin, and for Poisson evaluations where the normalization plays a role. A summary of all discussed likelihood formulations is shown in appendix \ref{appendix:summary}, table \ref{table:summary_formulas}.
 
Beyond the pure application in statistical analyses of data, there are several directions for further investigation.
On the mathematical side, it would be interesting to know if there exist further simplification for the standard multinomial formulas. On the statistical side, it might be interesting to study goodness-of-fit behavior or to allow the Prior parameter $\alpha$ to vary as a nuisance parameter. Finally, it would be interesting to study the new likelihood expressions in unsupervised or supervised machine learning models involving Monte Carlo estimates.

\section*{Acknowledgements}
We would like to thank Eberhard B\"ansch, Dmitry Chirkin and the anonymous referee for useful discussions and feedback.

\appendix

\section{Appendix}

\subsection{Overview of constructions and implementation}
\label{appendix:summary}
Table \ref{table:summary_formulas} summarizes all new formulas discussed in the paper. Python implementations and example usage can be found on \href{http://www.github.com/thoglu/mc\_uncertainty}{\texttt{http://www.github.com/thoglu/mc\_uncertainty}}.

\begin{table}
	\setlength\extrarowheight{2.5pt}
	\centering
	\begin{tabular}{c>{\centering\arraybackslash} m{0.33\textwidth} >{\centering\arraybackslash} m{0.33\textwidth}}
		\hline \hline
		\multicolumn{3}{c}{\text{\bfseries{Poisson likelihood }}}       \\ \hline
		\raisebox{-10pt}{"Infinite statistics"} & 
		\multicolumn{2}{|c}{\raisebox{-10pt}{$\displaystyle \prod_i \frac{ \mathrm{e}^{-\sum_j w_{j,i}} \cdot (\sum_j w_{j,i})^{k_i} } {k_i! }$} }   
		\\ \hline
		& \multicolumn{1}{|c}{standard form} & "product" form
		\\ \hline
		Equal weights & 
		\multicolumn{1}{|c}{\multirow{2}{*}{ eq. (\ref{eq:equal_form_one}) } } & \multirow{2}{*}{ eq. (\ref{eq:product_construction_equal}) }  \\ (or avg. weight per bin) & 	\multicolumn{2}{|c}{} \\ \hline
		\multirow{2}{*}{General weights} &    
		\multicolumn{1}{|c}{combinatorial (eq. \ref{eq:finite_poisson_combinatorial}) } & \multirow{2}{*}{ eq. (\ref{eq:product_construction_general}) } \\ & \multicolumn{1}{|c}{finite sum (eq. \ref{eq:finite_poisson_finitesum}) } &   \\ \hline  \hline  
		
		\multicolumn{3}{c}{	\text{\bfseries{Multinomial likelihood}}  }    \\ \hline
		\raisebox{-10pt}{"Infinite statistics"} &
		\multicolumn{2}{|c}{	\raisebox{-10pt}{$\displaystyle  k! \cdot \prod_{\mathrm{bins} \ i} \frac{1}{{k_i}!} \left(\frac{\sum_j w_{i,j}}{\sum_{u,\mathrm{all}} w_u}\right)^{k_i}$} }      \\ \hline
		& \multicolumn{1}{|c}{standard form} & "ratio" form\\\hline
		Equal weights & 
		\multicolumn{1}{|c}{\multirow{2}{*}{ eq. (\ref{eq:finite_multinomial_equal}) }} & \multirow{2}{*}{ eq. (\ref{eq:mn_ratio_form_finite}) } \\ (or avg. weight per bin) & \multicolumn{2}{|c}{} \\ \hline
		General weights &  \multicolumn{1}{|c}{eq. (\ref{eq:finite_mnomial_general})  }  
		&  eq. (\ref{eq:ratio_general})                \\ \hline \hline
		\multicolumn{3}{c}{	\text{\bfseries{Unbinned likelihood with MC-derived PDFs}}  }    \\ \hline
		\multicolumn{3}{c}{	\raisebox{-0pt}{ eq. (\ref{eq:mnomial_unbinned_relation}) - transform corresponding multinomial likelihood in categorical form ($k=1$) for each event }} \\\hline             
	\end{tabular}
	\caption{Summary of the different extended likelihood formulas for finite Monte Carlo statistics that are discussed in the paper. The equal-weight formulas are also applicable as approximations in the general case  assuming the average weight per bin. Implementations of all formulas can be found on \href{http://www.github.com/thoglu/mc\_uncertainty}{\texttt{http://www.github.com/thoglu/mc\_uncertainty}}. }\label{table:summary_formulas}
\end{table}

\subsection{Expectation of the Poisson factor under the gamma distribution}
The expectation values for the likelihood factors under the gamma distribution involve known definite integrals, see e.g. \cite{Bronshtein2013}.
For the Poisson likelihood, each expectation value evaluates to

\begin{align}
\mathrm{E}\left[\frac{{e^{-\lambda}}{\lambda}^{k} }{k!}\right]_{\mathrm{\textbf{G}}(\lambda;\alpha,\beta)} &=
\int_{0}^{\infty} \frac{e^{-\lambda} \cdot \lambda^k}{ k!} \cdot \frac{\beta^\alpha \cdot e^{-\beta \lambda} \cdot \lambda^{\alpha-1}}{ \Gamma(\alpha)} d \lambda \nonumber
\\
&=  \frac{\beta^{\alpha}}{ \Gamma(\alpha) \cdot k!} \cdot \int_{0}^{\infty} e^{-\lambda(1+\beta)} \cdot \lambda^{k+\alpha-1} d \lambda \nonumber\\ &= \frac{\beta^{\alpha}  \cdot \Gamma(k+\alpha)}{\Gamma(\alpha) \cdot k!\cdot(1+\beta)^{k+\alpha}}
\label{eq:exp_value_solution}
\end{align},
which is used in section \ref{sec:poisson}.

\subsection{Marginal likelihood for the multinomial case}
\label{appendix:marginal_llh_multinomial}
Here we describe the expanded calculation of the marginal likelihood (section \ref{sec:multinomial}, eq. \ref{eq:mnomial_scaled_dirichlet}), which is an integral over a multinomial factor and the scaled Dirichlet density. 

\begin{align}
&\begin{aligned}
\mathllap{L_\mathrm{\textbf{MN},finite, eq.}} &=\stackrel[\sum p_i\leq1]{}{\int_{p_1} \ldots \int_{p_{N-1}}} \mathrm{\textbf{MN}}(k_1, \ldots, k_{N}; p_1, \ldots p_N) \\ \cdot \frac{\Gamma(\alpha_{tot}^{*})}{\prod_i^{N} \Gamma(\alpha_i^{*}) } & \cdot  \left( \prod_i^{N} {\beta_i^{*}}^{\alpha_i^{*}} \right) \cdot \frac{ {p_1}^{\alpha_1^{*}-1} \ldots {p_{N-1}}^{\alpha_{N-1}^{*}-1} }{ \left(\beta_N^{*} (1-\sum_i^{N-1} p_i) + \sum_i^{N-1} \beta_i^{*} \cdot p_i\right)^{\alpha_{tot}^{*}}} \\ & \cdot \left(1-\sum_i^{N-1} {p_i}\right)^{\alpha_{N-1}^{*}} \ dp_1 \ldots dp_{N-1}
\end{aligned}\\
&\begin{aligned}
&= \frac{k!}{\prod_i^{N} k_i!}  \cdot \frac{\Gamma(k_{mc}^{*})}{\prod_i^{N} \Gamma(k_{mc,i}^{*}) } \cdot \left( \prod_i^{N} {\left(\frac{1}{w_i}\right)}^{k_{mc,i}^{*}} \right) \\ \cdot \stackrel[\sum p_i\leq1]{}{\int_{p_1} \ldots \int_{p_{N-1}}} &  \frac{p_1^{k_1+k_{mc,1}^{*}-1}  \ldots {p_{N-1}}^{k_{N-1}+k_{mc,N-1}^{*}-1}}{\left(1/w_N \cdot (1-\sum_i^{N-1} p_i)+\sum_i  p_i/w_i\right)^{k_{mc}^{*}}} \\ &\cdot \left(1-\sum_i^{N-1} {p_i}\right)^{k_N+k_{mc,N}^{*}-1} \ dp_1 \ldots dp_{N-1} 
\end{aligned}\\
&\begin{aligned}
&= \frac{k! \cdot \Gamma(k_{mc}^{*})}{\prod_i^{N} k_i! \cdot  \Gamma(k_{mc,i}^{*})}    \cdot \left( \prod_i^{N} {\left(\frac{w_N}{w_i}\right)}^{k_{mc,i}^{*}} \right) \\ \cdot \stackrel[\sum p_i\leq1]{}{\int_{p_1} \ldots \int_{p_{N-1}}} &  \frac{p_1^{k_1+k_{mc,1}^{*}-1}  \ldots {p_{N-1}}^{k_{N-1}+k_{mc,N-1}^{*}-1}}{  \left(1-\sum_i^{N-1}(1-w_N/w_i) \right)^{k_{mc}^{*}}} \\ &\cdot \left(1-\sum_i^{N-1} {p_i}\right)^{k_N+k_{mc,N}^{*}-1} \ dp_1 \ldots dp_{N-1} 
\end{aligned} \\
&\begin{aligned}
&= \frac{k! \Gamma(k_{mc}^{*})}{\Gamma(k+k_{mc}^{*})}  \prod_{i}^N \left( \frac{\Gamma(k_i+k_{mc,i}^{*})}{k_{i}!\cdot \Gamma(k_{mc,i}^{*})} \right)    \cdot \left( \prod_i^{N} {\left(\frac{w_N}{w_i}\right)}^{k_{mc,i}^{*}} \right) \\ \cdot \stackrel[\sum p_i\leq1]{}{\int_{p_1} \ldots \int_{p_{N-1}}} &  \frac{p_1^{k_1+k_{mc,1}^{*}-1}  \ldots {p_{N-1}}^{k_{N-1}+k_{mc,N-1}^{*}-1}}{  \left(1-\sum_i^{N-1}(1-w_N/w_i) \right)^{k_{mc}^{*}}} \\ &\frac{k! \cdot \Gamma(k_{mc}^{*})}{\prod_i^{N} \Gamma(k_i+k_{mc,i}^{*})}\cdot \left(1-\sum_i^{N-1} {p_i}\right)^{k_N+k_{mc,N}^{*}-1} \ dp_1 \ldots dp_{N-1} 
\end{aligned} \\
&\begin{aligned}
&= \mathrm{\textbf{DM}}(\bm{k}, \bm{k_{mc}^{*}}) \cdot \left( \prod_i^{N} {\left(\frac{w_N}{w_i}\right)}^{k_{mc,i}^*} \right) \cdot F_D(a;\bm{b};c,\bm{z})
\end{aligned}
\end{align}
with $a=k_{mc}^{*}$, $b_i=k_{mc,i}^{*}+k_i \ (i=1 \ldots N-1)$ , $c=k_{mc}^{*}+k$, $z_i=1-w_N/w_i \ (i=1 \ldots N-1)$. In the process, we manipulate the integrand following \cite{Laarhoven1988}, but take the expectation value of the full multinomial likelihood, instead of only one factor. In the last step, we exploit that the integral corresponds to an integral representation of $F_D(a;\bm{b};c,\bm{z})$ for $c>\sum b_i$.

\subsection{Series expansion for marginal Poisson likelihood with general weights}
\label{appendix:series_derivation}

Here we describe the follow-up calculation from section \ref{sec:poisson_general_weights} to derive a series-representation of the marginal Poisson likelihood for a single bin that can be calculated to arbitrary precision, and can be a useful tool as a crosscheck. The calculation gives

\begin{align}
L_{\mathrm{\textbf{P}, finite, gen.}} &=
 \mathrm{E}\left[\frac{{e^{-\lambda}}{\lambda}^{k} }{k!}\right]_{P_{\mathrm{general}}(\lambda)} \nonumber  \\
&=  \int \frac{{e^{-\lambda}}{\lambda}^{k} }{k!} \cdot C \cdot \sum_{l=0}^{\infty} {\delta_l \cdot \frac{ \lambda^{\rho+l-1} \cdot e^{-\lambda/w_N  }   }{ \Gamma(\rho+l) \cdot w_{\mathrm{N}}^{\rho+l} } } d \lambda  \nonumber  \\
& = C \sum_{l=0}^{\infty} \delta_l \cdot \frac{(1/w_N)^{k_{mc}^{*}+l}  \cdot \Gamma(k+l+k_{mc}^{*})}{\Gamma(l+k_{mc}^{*}) \cdot k_i!\cdot(1+1/w_N)^{k+l+k_{mc}^{*}}} \nonumber \\
&= C  \frac{(1/w_N)^{k_{mc}^{*}}  \cdot }{ k!\cdot(1+1/w_N)^{k+k_{mc}^{*}}}   \sum_{l=0}^{\infty} \delta_l \cdot \frac{ \Gamma(k+l+k_{mc}^{*}) \cdot (1/w_N)^{l} }{\Gamma(l+k_{mc}^{*}) \cdot (1+1/w_N)^{l}}
\label{eq:poisson_general_seriesexpansion}  
\end{align}, where eq. (\ref{eq:exp_value_solution}) has been used to solve the integral. The relevant parameters $\rho$, $C$ and $\delta_l$ are defined as described in section \ref{sec:poisson_general_weights}. The weight $w_N$ denotes the smallest of all weights in the bin. For equal weights, $\delta_l=0 \ \forall \ l>0$, $C=1$, and the formula reduces to eq. (\ref{eq:equal_form_one}).

\subsection{Mathematical identities}

\subsubsection{A combinatorial identity}

Using eq. (\ref{eq:multinomial_rep_for_fd}) and eq. (\ref{eq:fd_as_residue}) one can derive a generalization of the identity in eq. (\ref{eq:combinatorial_sum_complex}), namely

\begin{equation}
\stackrel[\sum k_i=K, \ k_i \geq 0]{}{\sum}{ \prod_i^{N}  {{m_i+k_i-1}\choose{k_i}} x_i^{k_i}} = \frac{1}{2 \pi i} \oint \frac{t^{M +K-1}}{\prod_i^{N} (t-x_i)^{m_i}}  dt \label{eq:contour_identity}
\end{equation}
with $M=\sum_i m_i$ and $K=\sum_i k_i$. It can also be derived following Egorychev's rules \cite{Egorychev1984} for combinatorial sums.

\subsubsection{Calculation of a few mathematical objects via the special function $R_n$}
\label{appendix:pgf_and_others}

The special function $R_n$ has been introduced by Carlson in the 60s \cite{Carlson1963} as a certain average under the Dirichlet distribution on the simplex. The results in section \ref{sec:poisson_general_weights} involve $R_n$ with $n>0$, which can be evaluated using eq. (\ref{eq:multinomial_rep_for_fd}) and eq. (\ref{eq:mathematical_newdef_lauricella}) via 
\begin{align}
R_n(\bm{b},\bm{z}) &=F_D(n+b; \bm{b_{-1}}; b; 1-z_{1}^{-1} \ldots 1-z_{N-1}^{-1}) \cdot \prod_i z_i^{-b_i} \\
&=\frac{\Gamma(n+1) \Gamma(b)}{\Gamma(b+n)} \cdot \frac{1}{2 \pi i} \oint \frac{t^{b+n-1}}{\prod_i^{N} (t-z_i)^{b_i}}  dt = \frac{\Gamma(n+1) \Gamma(b)}{\Gamma(b+n)} \cdot  D_n(\bm{b},\bm{z})  \label{eq:general_r_derivation}
\end{align}
where
\begin{align}
D_{n}= \frac{1}{n}\sum\limits_{k=1}^{n} \left[\left(\sum\limits_{i=1}^{N} b_{i} \cdot {z_{i}}^k \right) D_{n-k}\right]
\end{align} 
with $D_0=1$ and $N$ the number of distinct $b_i$ or $z_i$, i.e. the same as already defined in section \ref{sec:poisson_general_weights} with slightly simplified notation. In the following we describe some efficient calculations for a few additional mathematical objects that might be useful in statistical modeling and involve $R_n$ in some way or another.

\paragraph{Generalized Gamma-Poisson distribution:}

The main result in section \ref{sec:poisson_general_weights} involves the Poisson likelihood function. Here the function $R_n$ is involved in the generalization of the Gamma-Poisson mixture distribution and ultimately Poisson distribution. Using eq. (\ref{eq:multinomial_rep_for_fd}) we can re-phrase eq. (\ref{eq:finite_poisson_finitesum}) directly as a probability distribution via
\begin{align}
P_{gen}(k; \bm{\alpha},\bm{\beta}) &= \frac{\Gamma(\alpha+k)}{\Gamma(k+1)\Gamma(\alpha)} \cdot R_k(\bm{\alpha}, \frac{1}{1+\bm{\beta}}) \cdot \prod_i {\left( \frac{1}{(1+1/\beta_i)} \right) }^{\alpha_i} \\ 
&= D_k \left( \bm{\alpha}, \frac{1}{1+\bm{\beta}} \right) \cdot \prod_i {\left( \frac{1}{(1+1/\beta_i)} \right) }^{\alpha_i}
\end{align}
where $P_{gen}(k)$ is a probability distribution in $k$ and $\alpha_i$, $\beta_i$ can be thought of as shape and scale parameters from individual gamma distributions in the definition by convolutional factors (eq. \ref{eq:base_formula_with_convolution_of_gammas}). When all $\alpha_i$ are equal and all $\beta_i$ are equal, the formula reduces to the standard Gamma-Poisson mixture distribution, i.e. an integral over a Poisson and a single gamma distribution. 

\paragraph{Divided difference of a monomial:}

Using the definition of the divided difference operator as a contour integral \cite{book:divdiv}, and using eq. (\ref{eq:contour_identity}), we see that the divided difference of a monomial $t^{n+m-1}$, $[x_1, x_2, \dots, x_m; t^{n+m-1}]$ also involves $R_n$, and can be calculated via
\begin{align}
[x_1, x_2, \dots, x_m; t^{n+m-1}]&=\frac{1}{2 \pi i} \oint \frac{t^{n+m-1}}{\prod_i^{m} (t-x_i)}  dt \\
&= R_n(\bm{1}, \bm{x}) \cdot \frac{\Gamma(n+m)}{\Gamma(n+1) \Gamma(m)} = D_n(\bm{1}, \bm{x})
\end{align}
where $m$ is the positive number of points $x_i$, whose individual values can occur multiple times, and $n$ is a non-negative integer. 

\paragraph{Probability generating function of the Dirichlet-multinomial distribution:}

Looking at eq. (\ref{eq:contour_identity}) and eq. (\ref{eq:general_r_derivation}) we see that we can write the PGF of the Dirichlet-multinomial distribution, as
\begin{align}
\mathrm{PGF}_{\mathrm{DM}}& = \mathrm{E}_{\mathrm{DM}}\left[\prod_i z_i^{k_i}\right]=\stackrel[\sum_i k_i=k, \ k_i \geq 0]{}{\sum} \mathrm{\textbf{DM}}(\bm{k}; \bm{\alpha}) \prod_i z_i^{k_i} = R_k(\bm{\alpha},\bm{z}) \\
&=\frac{\Gamma(k+1) \Gamma(\alpha)}{\Gamma(\alpha+k)} \cdot D_k(\bm{\alpha},\bm{z}) 
\end{align}
where $\sum_i k_i = k$ and $\sum_i \alpha_i=\alpha$.
Now one could proceed and write down the characteristic function or moment-generating function of the Dirichlet-multinomial, respectively. The result of course also applies for the simpler beta-binomial distribution as a special case.

\paragraph{Moments of univariate B-Splines:}

It is also known that $R_n(\bm{m};\bm{x})$ with $n>0$ represents the $n$th raw moment $\mu_n$ of a normalized univariate B-spline with knot positions $x_i$ and knot multiplicities $m_i$ \cite{Carlson1991}. Similar to the PGF of the Dirichlet-multinomial distribution we can write
\begin{align}
\mu_n &= R_n(\bm{m};\bm{x}) = \int_{-\infty}^{\infty} t^n \cdot B(t|x_0 \dots x_k) dt \\
&= \frac{\Gamma(n+1) \Gamma(m)}{\Gamma(m+n)} \cdot D_n(\bm{m},\bm{x}) 
\end{align}
where $B(t)$ represents the B-spline, $k$ its order, and  $\sum_i m_i = m$. The result is a robust formula to calculate the $n$th B-Spline moment, even when knots overlap or are close to each other. One can also simply put all multiplicities to unity, i.e. $m_i=1$, and re-use the respective $x_i$ multiple times. 

\bibliographystyle{unsrt}
\bibliography{sampling_based_uncertainty}

\end{document}